\documentclass[prl,twocolumn,superscriptaddress,showpacs,groupedaddress]{revtex4}  

\usepackage{amssymb}
\usepackage{amsmath}
\usepackage{esint}
\usepackage{epsfig}
\usepackage{epstopdf}
\usepackage{bm}
\usepackage{graphicx,epsfig}
\usepackage{mathrsfs}
\usepackage{dcolumn}
\usepackage{color}
\usepackage{natbib}
\usepackage{CJK}
\usepackage{miniplot}
\usepackage{subfigure}
\usepackage{minitoc}
\usepackage{blindtext}
\usepackage[utf8]{inputenc}

\def\be{\begin{equation}}
\def\ee{\end{equation}}
\def\bea{\begin{eqnarray}}
\def\eea{\end{eqnarray}}
\def\la{\langle}
\def\ra{\rangle}

\allowdisplaybreaks[2]

\begin{document}

\title{Mesoscopic Superposition States Generated by Synthetic Spin-orbit Interaction in Fock-state Lattices}
\author{Da-Wei Wang}
\email{whatarewe@tamu.edu}
\affiliation{Texas A$\&$M University, College Station, TX 77843, USA}
\author{Han Cai}
\affiliation{Texas A$\&$M University, College Station, TX 77843, USA}
\author{Ren-Bao Liu}
\email{rbliu@phy.cuhk.edu.hk}
\affiliation{Department of Physics and Centre for Quantum Coherence, The Chinese University of Hong Kong, Hong Kong, China}
\author{Marlan O. Scully}
\affiliation{Texas A$\&$M University, College Station, TX 77843, USA}
\affiliation{Princeton University, Princeton, New Jersey 08544, USA}
\affiliation{Baylor University, Waco, TX 76706, USA}

\date{\today }

\begin{abstract}
Mesoscopic superposition states of photons can be prepared in three cavities interacting with the same two-level atom. By periodically modulating the three cavity frequencies around the transition frequency of the atom with $2\pi/3$ phase difference, the time reversal symmetry is broken and an optical circulator is generated with chiralities depending on the quantum state of the atom. A superposition of the atomic states can guide photons from one cavity to a mesoscopic superposition of the other two cavities. The physics can be understood in a finite spin-orbit-coupled Fock-state lattice where the atom and the cavities carry the spin and the orbit degrees of freedom, respectively. This scheme can be realized in circuit QED architectures and provides a new platform for exploring quantum information and topological physics in novel lattices.
\end{abstract}


\maketitle

Superposition of mesoscopically distinctive quantum states plays an important role in quantum metrology and lithography \cite{Boto2000,Israel2014,Wang2014}, quantum computation \cite{Ralph2003} and teleportation \cite{Enk2001}, and in the test of fundamental quantum theories \cite{Wenger2003,Greenberger1990}. Their many varieties, including the coherent state superposition (cat states) \cite{Yurke1986, Schleich1991,Brune1992,Monroe1996,Ourjoumtsev2007,Vlastakis2013}, NOON states \cite{Boto2000,Israel2014,Afek2010b}, entangled coherent states \cite{Enk2001,Sanders1992}, Greenberger-Horne-Zeilinger (GHZ) states \cite{Greenberger1990,Agarwal1997,Molmer1999,Monz2011}, and micro-macro entangled states \cite{Sekatski2012,Ghobadi2013}, are generally difficult to obtain due to their fragility and the requirement of large nonlinearities or post selection. The high-finesse \cite{Brune1996, Deleglise2008} and superconductor cavities \cite{Hofheinz2009} have enabled improved robustness and enhanced nonlinearity to such a level that the coherent state superposition of 100 photons has been realized \cite{Vlastakis2013}. By using the M\text{\o}lmer-S\text{\o}rensen approach \cite{Molmer1999}, GHZ states with 14 ions have been prepared \cite{Monz2011}. In contrast, scalable mechanisms are still rare to prepare NOON states, for which the highest number realized in experiments is 5 \cite{Afek2010b}.

High number Fock states can be systematically generated in superconducting resonators \cite{Guerlin2007,Sayrin2011} and circuit QED systems \cite{Hofheinz2008,Wang2008}. In particular, number states containing 15 photons \cite{Wang2008} have been prepared in superconductor circuits. Deterministic generation of NOON states up to $N=3$ was achieved by first creating entanglement between two resonators and then increasing their photon numbers \cite{Wang2011}. However, once entanglement is created, decoherence accompanies the remaining process, which hinders the scaling up to high NOON states. Therefore, it is favourable to first create high photon number states before generating entanglement. In this Letter, we realize the following transformation that can generate NOON states from number states,
\begin{equation}
a_0 \rightarrow a_1 |e\ra\la e|+a_2 |g\ra\la g|,
\label{opr}
\end{equation}
where $a_j$ ($j=0,1,2$) are the annihilation operators of three cavities, and $|e\ra$ and $|g\ra$ are the excited and ground states of a two-level atom. The photon state of a cavity is transferred to either one of two other cavities depending on the atomic states. 

The significance of this transformation is manifested in achieving various types of mesoscopic superposition states from number states or coherent states. We initially prepare an unentangled state 
\begin{equation}
|\psi\ra=\frac{1}{\sqrt{2}}|N,0,0\ra \left(|e\ra+|g\ra\right),
\end{equation}
where $|N,0,0\ra$ means that the first cavity contains $N$ photons while the other two cavities are in the vacuum state. After the transformation in Eq. (\ref{opr}), the state becomes
\begin{equation}
|\psi\ra\rightarrow\frac{1}{\sqrt{2}} \left(|0,N,0\ra|e\ra+|0,0,N\ra|g\ra\right),
\end{equation}
which is a micro-macro entangled state \cite{Sekatski2012,Ghobadi2013}. Applying a $\pi/2$ pulse to the above state yields
$
\left[\left(|0,N,0\ra-|0,0,N\ra\right)|e\ra+\left(|0,N,0\ra+|0,0,N\ra\right)|g\ra\right]/2,
$
in which the cavities are in different NOON states for different atomic states. If initially the first cavity is in a coherent state $|\alpha,0,0\ra$, the final photon states are $1/\sqrt{2}\left(|0,\alpha,0\ra\pm|0,0,\alpha\ra\right)$, i.e., entangled coherent states \cite{Sanders1992}.

\begin{figure*}[t]
    \epsfig{figure=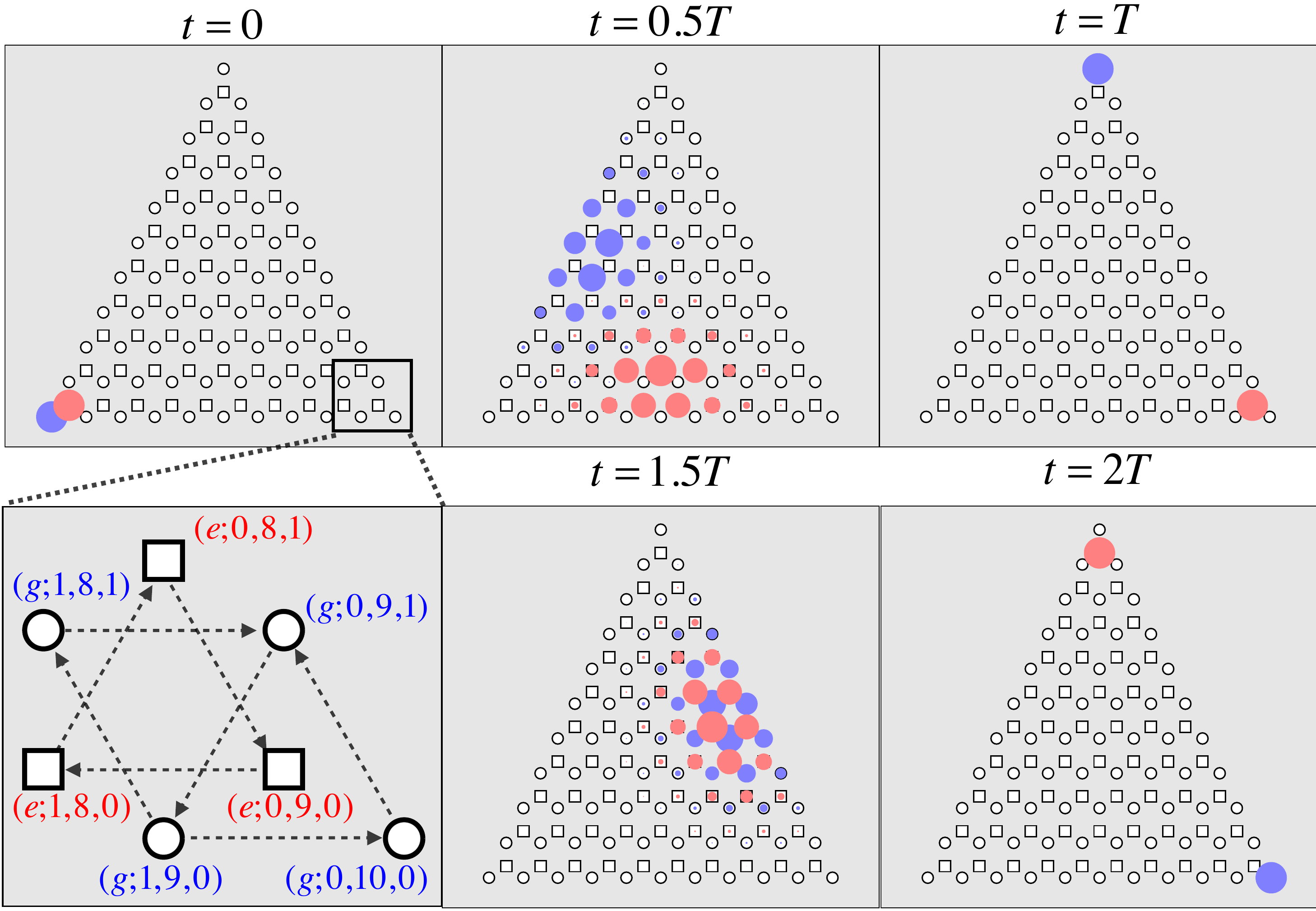, angle=0, width=0.68\textwidth}
\caption{(color online). {The dynamic evolution of a quantum state in the Fock-state lattice.} The empty circles (squares) denote the $|g\ra$ ($|e\ra$) sublattice. The radius of the blue (red) filled circles denote the probabilities (rescaled in each frame) of the quantum states on the $|g\ra$ ($|e\ra$) sublattice. The photon numbers of the states are as denoted in the enlarged figure. The arrows denote the transitions attached with the phase factor $-i$. Up and down triangles in the same sublattice have opposite fluxes. Up triangles in the two sublattices also have opposite fluxes. The lattice contains ten excitations, i.e., ${N}=10$. The initial state $|\psi_1\ra$ at $t=0$ is a superposition of $|g\ra |10,0,0\ra$ and $|e\ra |9,0,0\ra$, which travel in opposite directions on the two sublattices. }
\label{trans}
\end{figure*}

The transformation in Eq.(\ref{opr}) can be obtained by the following Hamiltonian
\begin{equation}
H=i\hbar\kappa\sigma_z\sum\limits_{j=0}^{2} a_{j+1}^\dagger a_j+h.c.,
\label{H}
\end{equation}
where the cavity index $j$ is cyclic, $\hbar$ is the reduced Plank constant, $\kappa$ is a real-number coupling constant, and $\sigma_z=|e\ra\la e|-|g\ra\la g|$ is the ${z}$-component of the pseudo-spin Pauli matrices of the atom. $H$ can be regarded as a spin-orbit-coupled Hamiltonian of a Fock-state lattice (Fig.~\ref{trans}), where the directions of photon currents depend on the pseudo-spin state of the atom, as will be shown later. The Hamiltonian can be diagonalized,
$
H=-2\hbar\kappa\sigma_z \sum_{j=0}^{2}\sin \left({2j\pi}/{3}\right)b_{j}^\dagger b_j,
$
with $b_j=1/\sqrt{3}\sum_{j^\prime=0}^2 \exp(ijj^\prime 2\pi/3)a_{j^\prime}$. The eigenfrequencies of the eigenmodes $b_0$, $b_1$ and $b_2$ are correspondingly $\omega_0=0$, $\omega_1=-\sqrt{3} \kappa\sigma_z$ and $\omega_2=\sqrt{3}\kappa\sigma_z$. The evolution of $a_0$ is 
\begin{equation}
\begin{aligned}
a_0\left(t\right)=&\frac{1}{\sqrt{3}}\sum\limits_{j=0}^2 e^{-i\omega_j t}b_j(0)\\
=&\frac{1}{3}\sum\limits_{j=0}^2\left[1+2\cos\left(\sqrt{3}\kappa\sigma_z t+\frac{2j\pi}{3}\right)\right]a_j(0),
\end{aligned}
\label{dyn}
\end{equation}
which yields $a_0(T)=a_1(0)|g\ra\la g|+a_2(0)|e\ra\la e|$ at $T\equiv 2\pi/3\sqrt{3}\kappa$. Similar equations of $a_1$ and $a_2$ lead to $a_0(0)=a_1(T)|e\ra\la e|+a_2(T)|g\ra\la g|$, i.e., Eq. (\ref{opr}).

One important feature of $H$ that realizes the above transformation rests in the complex coupling coefficient $i\kappa$, which introduces an effective magnetic field in the pseudo-lattice formed by the Fock states of the three cavities (as shown in Fig.\ref{trans} and discussed later). This synthetic magnetic field for photons breaks the time reversal symmetry and creates an optical circulator, which can be generated in circuit QED architectures \cite{Koch2010,Sliwa2015} and parametrically modulated coupled-resonators \cite{Estep2014}. Another key feature of our scheme is that the chirality of the circulator is opposite for $|e\ra$ and $|g\ra$ states due to the factor $\sigma_z$ in Eq.(\ref{H}). If $\kappa>0$, the ground state mode is $b_1$ for $|e\ra$ state and $b_2$ for $|g\ra$ state. $b_1$ and $b_2$ are photonic modes with opposite quasi-momenta, which drive the rotation in Eq.(\ref{dyn}). 

The Hamiltonian $H$ can be realized in three cavities with modulated frequencies $\nu_j=\nu+\Delta\sin(\nu_d t-2j\pi/3)$ coupled to the same two-level system. The interaction Hamiltonian under the rotating wave approximation is
\begin{equation}
H_I=\hbar\delta\sigma_z/2+\hbar g_v\sum\limits_{j=0}^{2}\left(\sigma^+a_je^{if\cos(\nu_dt-2j\pi/3)}+h.c.\right),
\label{Hi}
\end{equation}
where $\delta=\omega-\nu$ with $\omega$ being the atomic transition frequency, $g_v$ the vacuum Rabi frequency between the cavities and the atom, $\sigma^+=|e\ra\la g|$ the atomic raising operator, and $f=\Delta/\nu_d$. Under the condition $\nu_d\gg \sqrt{N}g_v$ with $N$ the total excitation number of the atom and photons, we obtain the effective Floquet Hamiltonian after adiabatically eliminating the fast oscillating terms \cite{Goldman2014,Jotzu2014a} as $H_I=H_0+H$ \cite{Supp}, where $H_0=\hbar\delta\sigma_z/2+\hbar g_v J_0(f)\sum_{j=0}^{2}(\sigma^+a_i+h.c.)$ and $H$ is defined in Eq.~(\ref{H}) with $\kappa= g_v^2\beta/\nu_d$ and $\beta=\sum_{n=1}^{\infty}2J^2_n(f)\sin(2n\pi/3)/n$. Here $J_n(f)$ is the $n$th order Bessel function of the first kind. When $\delta=0$ and $J_0(f)=0$ at $f=2.40$, we obtain $\beta\approx 0.307$ and $H_I=H$. Alternatively, one can also modulate the coupling strengths \cite{Liberato2007,Allman2014} between the cavities $j$ and the atom such as $g_j(t)=2g_v\cos(\nu_dt-2j\pi/3)$ to realize $\kappa=\sqrt{3}g^2_v/\nu_d$.

Helical currents exist in the lattice composed of the photon number states of the three cavities, as shown in Fig.\ref{trans}. The effective Hamiltonian $H_I=H_0+H$ conserves the total excitation number ${N}=\sum_{j=0}^{2}{n}_j+(\sigma_z+1)/2$ where ${n}_j=a_j^\dagger a_j$. The quantum states with constant ${N}$ form a finite triangular lattice. One of ${n}_j$'s is zero on each of the three triangular boundaries. This lattice has a similar structure to the Haldane model \cite{Haldane1988} with site-varying coupling coefficients. In particular, there are periodic magnetic fluxes, which are the key for topological insulators and helical edge states \cite{Kane2005}. The up and down triangles in the same sublattice have opposite effective magnetic fluxes. Due to the triangular boundaries, up triangles outnumber down triangles by the total number of photons and there are net fluxes in each of the whole sublattices, which are in particular obvious near the edges. The quantum states $|e\ra |N-1,0,0\ra$ and $|g\ra |N,0,0\ra$ travel near the edges in opposite directions, which results from the opposite net local effective magnetic field in the two sublattices.

We can understand the helical transportation from the dispersion relation in the eigen space of normal modes. The initial state is $|N,0,0\ra=a^{\dagger N}_0|\text{vac}\ra/\sqrt{N!}$ where $|\text{vac}\ra$ is the vacuum state. Since $a_0=1/\sqrt{3}\sum_{j=0}^2 b_j$, we expand $|N,0,0\ra$ in the basis of the normal modes $b_j$,
\begin{equation}
|N,0,0\ra=\sum\limits_{m_0,m_1,m_2}\sqrt{\frac{N!}{3^N m_0!m_1!m_2!}}|m_0,m_1,m_2\ra_b,
\end{equation}
where $|m_0,m_1,m_2\ra_b$ constrained by $\sum_{j=0}^2 m_j=N$ are the photon number states in $b_j$ modes. The factor $N!/m_1!m_2!m_3!$ reaches its maximum at $m_0\approx m_1\approx m_2\approx N/3$. The states with these photon numbers are concentrated near the corners and edges of the photon number lattice of $a_j$ modes. The energy and the quasi-momentum of the state $|m_0,m_1,m_2\ra_b$ are $E=\sqrt{3}\hbar\kappa\sigma_z (m_2-m_1)$ and $p=2\pi\hbar(m_2-m_1)/3$, where the direction in $a_0\rightarrow a_1\rightarrow a_2$ is defined as positive and the distance between two $a_j$ modes is one. The group velocity is therefore
\begin{equation}
v_g=\frac{\partial E}{\partial p}=\frac{3\sqrt{3}\kappa\sigma_z}{2\pi}=\frac{\sigma_z}{T}.
\end{equation}
After time $T$, the photons are transported from mode $a_0$ to $a_1$ when $\sigma_z=1$ and to $a_2$ when $\sigma_z=-1$, which is consistent with the conclusion of Eq.(\ref{dyn}).

In our previous study on topological superradiance lattices \cite{Wang2015b}, plane wave modes are coupled with extended ensemble of atoms and the superradiance momentum states of atoms form the approximately infinite lattice structures. Here we have a single atom coupled with cavity modes. Cavity photon number states form the lattice structures, which contain edges determined by quantum electrodynamics (no negative photon number states exist). In this finite lattice, the hopping rates depend on the sites, which is an inherit property of the annihilation operators. However, the helical currents are robust to the site-varying coupling strength. Moreover, it results in the synchronized and nondispersive ($\partial E/\partial p=E/p$ and $\partial^2E/\partial p^2=0$) transportation of different number states. At $t=mT$ with $m$ an integer, the excitation concentrates on one lattice site at one of the three corners, while in a lattice with homogeneous coupling strengths, the excitation is scattered by the corners and distributed all over the lattice for large $t$ \cite{Supp}.

\begin{figure}[t]
    \epsfig{figure=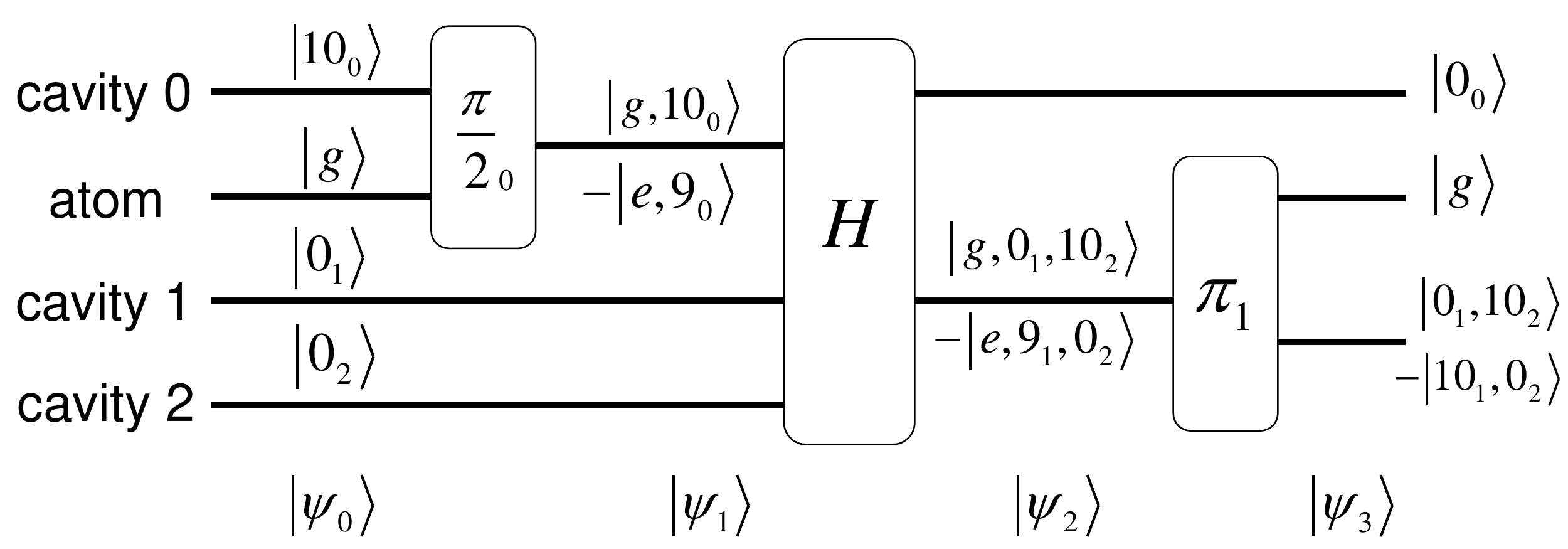, angle=0, width=0.48\textwidth}
\caption{ {Quantum circuit of the scheme preparing NOON states.} The scheme is valid for arbitrary ${N}$ and we take ${N}=10$ here. Each line represents an inseparable quantum state. $n_j$ is the $n$ Fock state of the cavity mode $j$. The blocks $\theta_j$ represent the $\theta$ Rabi rotation of the atom interacting with the cavity $j$. $H$ represents the evolution with the interaction Hamiltonian in Eq.(\ref{H}) for time $T$. $|\psi_i\ra$ is the direct product of the quantum states at each stage. We omit the normalization factors for superposition states.}
\label{circuit}
\end{figure}

NOON states can be realized without being entangled with the atom at the end, as shown in Fig.\ref{circuit}. We first prepare the state $|\psi_0\ra=|g\ra|N,0,0\ra$ and all the cavities but $a_0$ are initially out of resonance.~After a $\pi/2$ Rabi rotation, $|\psi_1\ra=1/\sqrt{2}\left(|g\ra|N,0,0\ra-|e\ra|N-1,0,0\ra\right)$.~Then we modulate the cavity frequencies and after time $T$, $|\psi_2\ra=1/\sqrt{2}\left(|g\ra|0,0,N\ra-|e\ra|0,N-1,0\ra\right)$.~We then tune all cavities but $a_1$ out of resonance.~After a $\pi$ Rabi rotation, the final state is $|\psi_3\ra=1/\sqrt{2}|g\ra\left(|0,0,N\ra-|0,N,0\ra\right)$.~If we replace the number state $|N,0,0\ra$ with the coherent state $|\alpha,0,0\ra$, we can prepare the entangled coherent state $1/\sqrt{2}(|0,0,\alpha\ra-|0,\alpha,0\ra)$ with small discrepancies due to the different Rabi frequencies of different number states, which can be neglected when $\alpha$ is large (see \cite{Supp} for the discrepancy and the cat state preparation).

The photonic GHZ states can also be prepared by making a chain of three-cavity systems. By initially preparing a quantum state $1/\sqrt{2}(|e\ra+|g\ra)\otimes_{j=1}^M|N,0,0\ra_j$ and sending the atom through a chain that contains $M$ three-cavity systems, the final state is $1/\sqrt{2}(|e\ra \otimes_{j=1}^M|0,N,0\ra_j+|g\ra\otimes_{j=1}^M|0,0,N\ra_j)$, which is the GHZ state for $N=1$ after a $\pi/2$ Rabi rotation of the atom. More interestingly, for $N\gg 1$, highly entangled GHZ-NOON states are prepared.

Mesoscopic superposition states can be prepared even if we only have two cavities \cite{Supp}, $H=\hbar\kappa\sigma_z (ia_{0}^\dagger a_1-ia_{1}^\dagger a_0)=\kappa\sigma_z J_y$ where we have used the Schwinger boson representation of the angular momentum $J_y=\hbar(ia_{0}^\dagger a_1-ia_{1}^\dagger a_0)$. This Hamiltonian is equivalent to that of Faraday rotation \cite{Zou2002,Leuenberger2006} and the quantum optical Fredkin gate \cite{Gerry2001,Sharypov2013}. The large spin of Schwinger bosons (photons in the two cavity modes) rotates in opposite directions conditioned by the small spin states. The state $|J_z=-N\ra\left(|e\ra+|g\ra\right)/\sqrt{2}$ can evolve to $\left(|J_x=-N\ra|e\ra+e^{i\phi}|J_x=N\ra|g\ra\right)/\sqrt{2}$ with $\phi$ a phase angle. However $|J_x=\pm N\ra$ involve superposition of two cavity modes and hinder the extraction of the final states. Macroscopic rotation up to six degrees has been realized in quantum dot cavities \cite{Arnold2015}, which nevertheless cannot be used to prepare NOON states. Besides, there is no synthetic magnetic field for only two modes.

\begin{figure}[t]
    \epsfig{figure=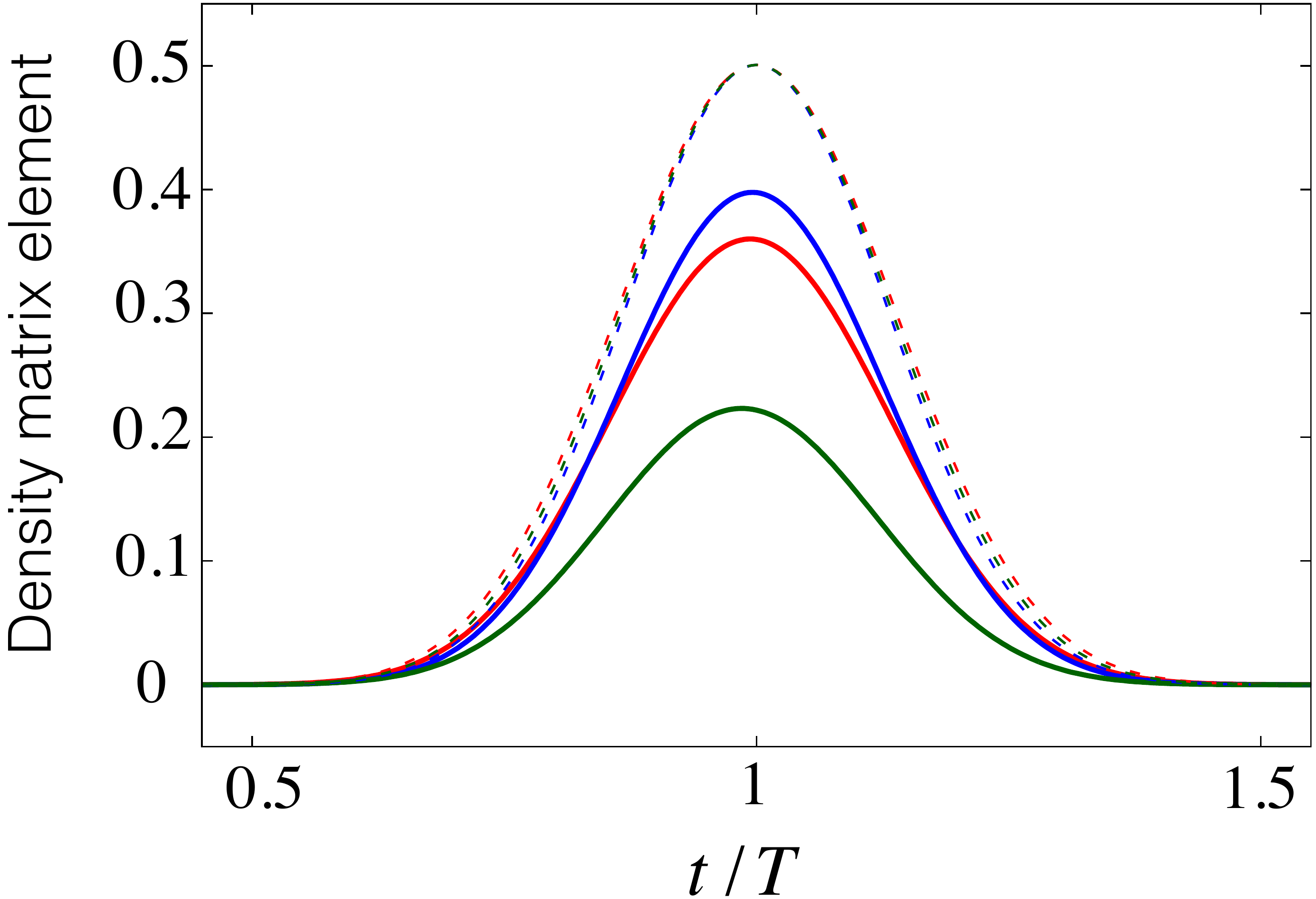, angle=0, width=0.45\textwidth}\\
\caption{(color online). Evolution of the density matrix elements with (solid lines) and without (dashed lines) dissipation, $\left\langle e; 0,9,0|\rho(t)|e;0,9,0\right\rangle$ (red lines), $\left \langle g; 0,0,10|\rho(t)|g;0,0,10\right\rangle$ (blue lines) and $|\left\langle g; 0,0,10|\rho(t)|e;0,9,0\right\rangle|$ (green lines). The relaxation and dephasing times of the two-level system are 650 ns and 150 ns, respectively. The cavity relaxation time is 3.47 $\mu$s. $T=80$ ns.}
\label{dissipation}
\end{figure}

Next we analyse the experimental feasibility and the robustness of this scheme against noises. High number Fock states can be prepared on demand in various cavity systems \cite{Guerlin2007,Sayrin2011,Hofheinz2008,Wang2008,Krause1987,Bertet2002}. In particular, we can couple three superconductor resonators to the same superconducting phase qubit, where Fock states up to photon number 15 can be achieved \cite{Wang2008}. The qubit relaxation and dephasing times are typically tens of microseconds \cite{Rigetti2012,Jeffrey2014,Cleland2015}. The coupling strength between the qubit and the resonators can be hundreds of MHz \cite{Niemczyk2010,Forn2010}. The single photon relaxation time of fixed frequency resonators can be milliseconds \cite{Cleland2015,Reagor2015}. Tunable resonators based on kinetic inductance can also have a relaxation time as long as 6 $\mu$s \cite{Vissers2015}. The time modulation of the resonator frequencies can be realized by tuning a superconducting quantum interference device coupled to the resonators as those in the dynamic Casimir effect \cite{Laloy2008,Sandberg2008,Johansson2009,Wilson2011} or by permittivity modulation in parametrically modulated resonators \cite{Estep2014}. We adopt parameters that are typically achievable in superconductor circuit experiments \cite{Hofheinz2008, Wang2011} and simulate the same process as in Fig.\ref{trans} but with dissipation. The evolution of the density matrix components on the target state $\left\langle \psi_2|\rho(t)|\psi_2\right\rangle$ is plotted in Fig.\ref{dissipation}. It is clear that at time $t=T$, the density matrix has a major overlap with the target state $|\psi_2\ra$. 

In conclusion, mesoscopic superposition states of photons can be prepared via the synthetic spin-orbit interaction of three cavities coupled to the same two-level system. The mechanism can be readily realized in circuit QED systems and provide non-magnetic chiral building blocks. The photon number lattices are similar to the Haldane model with site-dependent coupling strengths and triangular boundaries.~Besides applications in quantum metrology and quantum information, this study provides a new platform for investigating topological properties in novel lattices.

The authors thank A. N. Cleland, H. Dong, W. Ge, S. M. Girvin, D. R. Herschbach and L. Wang for helpful discussion. We gratefully acknowledge the support of the National Science Foundation Grants No. PHY-1241032 (INSPIRE CREATIV) and the Robert A. Welch Foundation (Grant No. A-1261). H. Cai is supported by Herman F. Heep and Minnie Belle Heep Texas A$\&$M University Endowed Fund. R.-B.Liu was supported by Hong Kong RGC/GRF Project 401413 and CUHK VC's One-Off Discretionary Fund.\\

\bibliographystyle{apsrev4-1}

\bibliography{sl}

\begin{thebibliography}{57}%
\makeatletter
\providecommand \@ifxundefined [1]{%
 \@ifx{#1\undefined}
}%
\providecommand \@ifnum [1]{%
 \ifnum #1\expandafter \@firstoftwo
 \else \expandafter \@secondoftwo
 \fi
}%
\providecommand \@ifx [1]{%
 \ifx #1\expandafter \@firstoftwo
 \else \expandafter \@secondoftwo
 \fi
}%
\providecommand \natexlab [1]{#1}%
\providecommand \enquote  [1]{``#1''}%
\providecommand \bibnamefont  [1]{#1}%
\providecommand \bibfnamefont [1]{#1}%
\providecommand \citenamefont [1]{#1}%
\providecommand \href@noop [0]{\@secondoftwo}%
\providecommand \href [0]{\begingroup \@sanitize@url \@href}%
\providecommand \@href[1]{\@@startlink{#1}\@@href}%
\providecommand \@@href[1]{\endgroup#1\@@endlink}%
\providecommand \@sanitize@url [0]{\catcode `\\12\catcode `\$12\catcode
  `\&12\catcode `\#12\catcode `\^12\catcode `\_12\catcode `\%12\relax}%
\providecommand \@@startlink[1]{}%
\providecommand \@@endlink[0]{}%
\providecommand \url  [0]{\begingroup\@sanitize@url \@url }%
\providecommand \@url [1]{\endgroup\@href {#1}{\urlprefix }}%
\providecommand \urlprefix  [0]{URL }%
\providecommand \Eprint [0]{\href }%
\providecommand \doibase [0]{http://dx.doi.org/}%
\providecommand \selectlanguage [0]{\@gobble}%
\providecommand \bibinfo  [0]{\@secondoftwo}%
\providecommand \bibfield  [0]{\@secondoftwo}%
\providecommand \translation [1]{[#1]}%
\providecommand \BibitemOpen [0]{}%
\providecommand \bibitemStop [0]{}%
\providecommand \bibitemNoStop [0]{.\EOS\space}%
\providecommand \EOS [0]{\spacefactor3000\relax}%
\providecommand \BibitemShut  [1]{\csname bibitem#1\endcsname}%
\let\auto@bib@innerbib\@empty
\bibitem [{\citenamefont {Boto}\ \emph {et~al.}(2000)\citenamefont {Boto},
  \citenamefont {Kok}, \citenamefont {Abrams}, \citenamefont {Braunstein},
  \citenamefont {Williams},\ and\ \citenamefont {Dowling}}]{Boto2000}%
  \BibitemOpen
  \bibfield  {author} {\bibinfo {author} {\bibfnamefont {A.~N.}\ \bibnamefont
  {Boto}}, \bibinfo {author} {\bibfnamefont {P.}~\bibnamefont {Kok}}, \bibinfo
  {author} {\bibfnamefont {D.~S.}\ \bibnamefont {Abrams}}, \bibinfo {author}
  {\bibfnamefont {S.~L.}\ \bibnamefont {Braunstein}}, \bibinfo {author}
  {\bibfnamefont {C.~P.}\ \bibnamefont {Williams}}, \ and\ \bibinfo {author}
  {\bibfnamefont {J.~P.}\ \bibnamefont {Dowling}},\ }\href@noop {} {\bibfield
  {journal} {\bibinfo  {journal} {Physical Review Letters}\ }\textbf {\bibinfo
  {volume} {85}},\ \bibinfo {pages} {2733} (\bibinfo {year}
  {2000})}\BibitemShut {NoStop}%
\bibitem [{\citenamefont {Israel}\ \emph {et~al.}(2014)\citenamefont {Israel},
  \citenamefont {Rosen},\ and\ \citenamefont {Silberberg}}]{Israel2014}%
  \BibitemOpen
  \bibfield  {author} {\bibinfo {author} {\bibfnamefont {Y.}~\bibnamefont
  {Israel}}, \bibinfo {author} {\bibfnamefont {S.}~\bibnamefont {Rosen}}, \
  and\ \bibinfo {author} {\bibfnamefont {Y.}~\bibnamefont {Silberberg}},\
  }\href@noop {} {\bibfield  {journal} {\bibinfo  {journal} {Physical Review
  Letters}\ }\textbf {\bibinfo {volume} {112}},\ \bibinfo {pages} {103604}
  (\bibinfo {year} {2014})}\BibitemShut {NoStop}%
\bibitem [{\citenamefont {Wang}\ and\ \citenamefont {Scully}(2014)}]{Wang2014}%
  \BibitemOpen
  \bibfield  {author} {\bibinfo {author} {\bibfnamefont {D.-W.}\ \bibnamefont
  {Wang}}\ and\ \bibinfo {author} {\bibfnamefont {M.~O.}\ \bibnamefont
  {Scully}},\ }\href@noop {} {\bibfield  {journal} {\bibinfo  {journal}
  {Physical Review Letters}\ }\textbf {\bibinfo {volume} {113}},\ \bibinfo
  {pages} {083601} (\bibinfo {year} {2014})}\BibitemShut {NoStop}%
\bibitem [{\citenamefont {Ralph}\ \emph {et~al.}(2003)\citenamefont {Ralph},
  \citenamefont {Gilchrist}, \citenamefont {Milburn}, \citenamefont {Munro},\
  and\ \citenamefont {Glancy}}]{Ralph2003}%
  \BibitemOpen
  \bibfield  {author} {\bibinfo {author} {\bibfnamefont {T.~C.}\ \bibnamefont
  {Ralph}}, \bibinfo {author} {\bibfnamefont {A.}~\bibnamefont {Gilchrist}},
  \bibinfo {author} {\bibfnamefont {G.~J.}\ \bibnamefont {Milburn}}, \bibinfo
  {author} {\bibfnamefont {W.~J.}\ \bibnamefont {Munro}}, \ and\ \bibinfo
  {author} {\bibfnamefont {S.}~\bibnamefont {Glancy}},\ }\href@noop {}
  {\bibfield  {journal} {\bibinfo  {journal} {Physical Review A}\ }\textbf
  {\bibinfo {volume} {68}},\ \bibinfo {pages} {042319} (\bibinfo {year}
  {2003})}\BibitemShut {NoStop}%
\bibitem [{\citenamefont {van Enk}\ and\ \citenamefont
  {Hirota}(2001)}]{Enk2001}%
  \BibitemOpen
  \bibfield  {author} {\bibinfo {author} {\bibfnamefont {S.~J.}\ \bibnamefont
  {van Enk}}\ and\ \bibinfo {author} {\bibfnamefont {O.}~\bibnamefont
  {Hirota}},\ }\href@noop {} {\bibfield  {journal} {\bibinfo  {journal}
  {Physical Review A}\ }\textbf {\bibinfo {volume} {64}},\ \bibinfo {pages}
  {022313} (\bibinfo {year} {2001})}\BibitemShut {NoStop}%
\bibitem [{\citenamefont {Wenger}\ \emph {et~al.}(2003)\citenamefont {Wenger},
  \citenamefont {Hafezi}, \citenamefont {Grosshans}, \citenamefont
  {Tualle-Brouri},\ and\ \citenamefont {Grangier}}]{Wenger2003}%
  \BibitemOpen
  \bibfield  {author} {\bibinfo {author} {\bibfnamefont {J.}~\bibnamefont
  {Wenger}}, \bibinfo {author} {\bibfnamefont {M.}~\bibnamefont {Hafezi}},
  \bibinfo {author} {\bibfnamefont {F.}~\bibnamefont {Grosshans}}, \bibinfo
  {author} {\bibfnamefont {R.}~\bibnamefont {Tualle-Brouri}}, \ and\ \bibinfo
  {author} {\bibfnamefont {P.}~\bibnamefont {Grangier}},\ }\href@noop {}
  {\bibfield  {journal} {\bibinfo  {journal} {Physical Review A}\ }\textbf
  {\bibinfo {volume} {67}},\ \bibinfo {pages} {012105} (\bibinfo {year}
  {2003})}\BibitemShut {NoStop}%
\bibitem [{\citenamefont {Greenberger}\ \emph {et~al.}(1990)\citenamefont
  {Greenberger}, \citenamefont {Horne}, \citenamefont {Shimony},\ and\
  \citenamefont {Zeilinger}}]{Greenberger1990}%
  \BibitemOpen
  \bibfield  {author} {\bibinfo {author} {\bibfnamefont {D.~M.}\ \bibnamefont
  {Greenberger}}, \bibinfo {author} {\bibfnamefont {M.~A.}\ \bibnamefont
  {Horne}}, \bibinfo {author} {\bibfnamefont {A.}~\bibnamefont {Shimony}}, \
  and\ \bibinfo {author} {\bibfnamefont {A.}~\bibnamefont {Zeilinger}},\
  }\href@noop {} {\bibfield  {journal} {\bibinfo  {journal} {American Journal
  of Physics}\ }\textbf {\bibinfo {volume} {58}},\ \bibinfo {pages} {1131}
  (\bibinfo {year} {1990})}\BibitemShut {NoStop}%
\bibitem [{\citenamefont {Yurke}\ and\ \citenamefont
  {Stoler}(1986)}]{Yurke1986}%
  \BibitemOpen
  \bibfield  {author} {\bibinfo {author} {\bibfnamefont {B.}~\bibnamefont
  {Yurke}}\ and\ \bibinfo {author} {\bibfnamefont {D.}~\bibnamefont {Stoler}},\
  }\href@noop {} {\bibfield  {journal} {\bibinfo  {journal} {Physical Review
  Letters}\ }\textbf {\bibinfo {volume} {57}},\ \bibinfo {pages} {13} (\bibinfo
  {year} {1986})}\BibitemShut {NoStop}%
\bibitem [{\citenamefont {Schleich}\ \emph {et~al.}(1991)\citenamefont
  {Schleich}, \citenamefont {Pernigo},\ and\ \citenamefont
  {Kien}}]{Schleich1991}%
  \BibitemOpen
  \bibfield  {author} {\bibinfo {author} {\bibfnamefont {W.}~\bibnamefont
  {Schleich}}, \bibinfo {author} {\bibfnamefont {M.}~\bibnamefont {Pernigo}}, \
  and\ \bibinfo {author} {\bibfnamefont {F.~L.}\ \bibnamefont {Kien}},\
  }\href@noop {} {\bibfield  {journal} {\bibinfo  {journal} {Physical Review
  A}\ }\textbf {\bibinfo {volume} {44}},\ \bibinfo {pages} {2172} (\bibinfo
  {year} {1991})}\BibitemShut {NoStop}%
\bibitem [{\citenamefont {Brune}\ \emph {et~al.}(1992)\citenamefont {Brune},
  \citenamefont {Haroche}, \citenamefont {Raimond}, \citenamefont
  {Davidovich},\ and\ \citenamefont {Zagury}}]{Brune1992}%
  \BibitemOpen
  \bibfield  {author} {\bibinfo {author} {\bibfnamefont {M.}~\bibnamefont
  {Brune}}, \bibinfo {author} {\bibfnamefont {S.}~\bibnamefont {Haroche}},
  \bibinfo {author} {\bibfnamefont {J.~M.}\ \bibnamefont {Raimond}}, \bibinfo
  {author} {\bibfnamefont {L.}~\bibnamefont {Davidovich}}, \ and\ \bibinfo
  {author} {\bibfnamefont {N.}~\bibnamefont {Zagury}},\ }\href@noop {}
  {\bibfield  {journal} {\bibinfo  {journal} {Physical Review A}\ }\textbf
  {\bibinfo {volume} {45}},\ \bibinfo {pages} {5193} (\bibinfo {year}
  {1992})}\BibitemShut {NoStop}%
\bibitem [{\citenamefont {Monroe}\ \emph {et~al.}(1996)\citenamefont {Monroe},
  \citenamefont {Meekhof}, \citenamefont {King},\ and\ \citenamefont
  {Wineland}}]{Monroe1996}%
  \BibitemOpen
  \bibfield  {author} {\bibinfo {author} {\bibfnamefont {C.}~\bibnamefont
  {Monroe}}, \bibinfo {author} {\bibfnamefont {D.~M.}\ \bibnamefont {Meekhof}},
  \bibinfo {author} {\bibfnamefont {B.~E.}\ \bibnamefont {King}}, \ and\
  \bibinfo {author} {\bibfnamefont {D.~J.}\ \bibnamefont {Wineland}},\
  }\href@noop {} {\bibfield  {journal} {\bibinfo  {journal} {Science}\ }\textbf
  {\bibinfo {volume} {272}},\ \bibinfo {pages} {1131} (\bibinfo {year}
  {1996})}\BibitemShut {NoStop}%
\bibitem [{\citenamefont {Ourjoumtsev}\ \emph {et~al.}(2007)\citenamefont
  {Ourjoumtsev}, \citenamefont {Jeong}, \citenamefont {Tualle-Brouri},\ and\
  \citenamefont {Grangier}}]{Ourjoumtsev2007}%
  \BibitemOpen
  \bibfield  {author} {\bibinfo {author} {\bibfnamefont {A.}~\bibnamefont
  {Ourjoumtsev}}, \bibinfo {author} {\bibfnamefont {H.}~\bibnamefont {Jeong}},
  \bibinfo {author} {\bibfnamefont {R.}~\bibnamefont {Tualle-Brouri}}, \ and\
  \bibinfo {author} {\bibfnamefont {P.}~\bibnamefont {Grangier}},\ }\href@noop
  {} {\bibfield  {journal} {\bibinfo  {journal} {Nature}\ }\textbf {\bibinfo
  {volume} {448}},\ \bibinfo {pages} {784} (\bibinfo {year}
  {2007})}\BibitemShut {NoStop}%
\bibitem [{\citenamefont {Vlastakis}\ \emph {et~al.}(2013)\citenamefont
  {Vlastakis}, \citenamefont {Kirchmair}, \citenamefont {Leghtas},
  \citenamefont {Nigg}, \citenamefont {Frunzio}, \citenamefont {Girvin},
  \citenamefont {Mirrahimi}, \citenamefont {Devoret},\ and\ \citenamefont
  {Schoelkopf}}]{Vlastakis2013}%
  \BibitemOpen
  \bibfield  {author} {\bibinfo {author} {\bibfnamefont {B.}~\bibnamefont
  {Vlastakis}}, \bibinfo {author} {\bibfnamefont {G.}~\bibnamefont
  {Kirchmair}}, \bibinfo {author} {\bibfnamefont {Z.}~\bibnamefont {Leghtas}},
  \bibinfo {author} {\bibfnamefont {S.~E.}\ \bibnamefont {Nigg}}, \bibinfo
  {author} {\bibfnamefont {L.}~\bibnamefont {Frunzio}}, \bibinfo {author}
  {\bibfnamefont {S.~M.}\ \bibnamefont {Girvin}}, \bibinfo {author}
  {\bibfnamefont {M.}~\bibnamefont {Mirrahimi}}, \bibinfo {author}
  {\bibfnamefont {M.~H.}\ \bibnamefont {Devoret}}, \ and\ \bibinfo {author}
  {\bibfnamefont {R.~J.}\ \bibnamefont {Schoelkopf}},\ }\href@noop {}
  {\bibfield  {journal} {\bibinfo  {journal} {Science}\ }\textbf {\bibinfo
  {volume} {342}},\ \bibinfo {pages} {607} (\bibinfo {year}
  {2013})}\BibitemShut {NoStop}%
\bibitem [{\citenamefont {Afek}\ \emph {et~al.}(2010)\citenamefont {Afek},
  \citenamefont {Ambar},\ and\ \citenamefont {Silberberg}}]{Afek2010b}%
  \BibitemOpen
  \bibfield  {author} {\bibinfo {author} {\bibfnamefont {I.}~\bibnamefont
  {Afek}}, \bibinfo {author} {\bibfnamefont {O.}~\bibnamefont {Ambar}}, \ and\
  \bibinfo {author} {\bibfnamefont {Y.}~\bibnamefont {Silberberg}},\
  }\href@noop {} {\bibfield  {journal} {\bibinfo  {journal} {Science}\ }\textbf
  {\bibinfo {volume} {328}},\ \bibinfo {pages} {879} (\bibinfo {year}
  {2010})}\BibitemShut {NoStop}%
\bibitem [{\citenamefont {Sanders}(1992)}]{Sanders1992}%
  \BibitemOpen
  \bibfield  {author} {\bibinfo {author} {\bibfnamefont {B.~C.}\ \bibnamefont
  {Sanders}},\ }\href@noop {} {\bibfield  {journal} {\bibinfo  {journal}
  {Physical Review A}\ }\textbf {\bibinfo {volume} {45}},\ \bibinfo {pages}
  {6811} (\bibinfo {year} {1992})}\BibitemShut {NoStop}%
\bibitem [{\citenamefont {Agarwal}\ \emph {et~al.}(1997)\citenamefont
  {Agarwal}, \citenamefont {Puri},\ and\ \citenamefont {Singh}}]{Agarwal1997}%
  \BibitemOpen
  \bibfield  {author} {\bibinfo {author} {\bibfnamefont {G.~S.}\ \bibnamefont
  {Agarwal}}, \bibinfo {author} {\bibfnamefont {R.~R.}\ \bibnamefont {Puri}}, \
  and\ \bibinfo {author} {\bibfnamefont {R.~P.}\ \bibnamefont {Singh}},\
  }\href@noop {} {\bibfield  {journal} {\bibinfo  {journal} {Physical Review
  A}\ }\textbf {\bibinfo {volume} {56}},\ \bibinfo {pages} {2249} (\bibinfo
  {year} {1997})}\BibitemShut {NoStop}%
\bibitem [{\citenamefont {Mølmer}\ and\ \citenamefont
  {Sørensen}(1999)}]{Molmer1999}%
  \BibitemOpen
  \bibfield  {author} {\bibinfo {author} {\bibfnamefont {K.}~\bibnamefont
  {Mølmer}}\ and\ \bibinfo {author} {\bibfnamefont {A.}~\bibnamefont
  {Sørensen}},\ }\href@noop {} {\bibfield  {journal} {\bibinfo  {journal}
  {Physical Review Letters}\ }\textbf {\bibinfo {volume} {82}},\ \bibinfo
  {pages} {1835} (\bibinfo {year} {1999})}\BibitemShut {NoStop}%
\bibitem [{\citenamefont {Monz}\ \emph {et~al.}(2011)\citenamefont {Monz},
  \citenamefont {Schindler}, \citenamefont {Barreiro}, \citenamefont {Chwalla},
  \citenamefont {Nigg}, \citenamefont {Coish}, \citenamefont {Harlander},
  \citenamefont {Hansel}, \citenamefont {Hennrich},\ and\ \citenamefont
  {Blatt}}]{Monz2011}%
  \BibitemOpen
  \bibfield  {author} {\bibinfo {author} {\bibfnamefont {T.}~\bibnamefont
  {Monz}}, \bibinfo {author} {\bibfnamefont {P.}~\bibnamefont {Schindler}},
  \bibinfo {author} {\bibfnamefont {J.~T.}\ \bibnamefont {Barreiro}}, \bibinfo
  {author} {\bibfnamefont {M.}~\bibnamefont {Chwalla}}, \bibinfo {author}
  {\bibfnamefont {D.}~\bibnamefont {Nigg}}, \bibinfo {author} {\bibfnamefont
  {W.~A.}\ \bibnamefont {Coish}}, \bibinfo {author} {\bibfnamefont
  {M.}~\bibnamefont {Harlander}}, \bibinfo {author} {\bibfnamefont
  {W.}~\bibnamefont {Hansel}}, \bibinfo {author} {\bibfnamefont
  {M.}~\bibnamefont {Hennrich}}, \ and\ \bibinfo {author} {\bibfnamefont
  {R.}~\bibnamefont {Blatt}},\ }\href@noop {} {\bibfield  {journal} {\bibinfo
  {journal} {Physical Review Letters}\ }\textbf {\bibinfo {volume} {106}},\
  \bibinfo {pages} {130506} (\bibinfo {year} {2011})}\BibitemShut {NoStop}%
\bibitem [{\citenamefont {Sekatski}\ \emph {et~al.}(2012)\citenamefont
  {Sekatski}, \citenamefont {Sangouard}, \citenamefont {Stobińska},
  \citenamefont {Bussières}, \citenamefont {Afzelius},\ and\ \citenamefont
  {Gisin}}]{Sekatski2012}%
  \BibitemOpen
  \bibfield  {author} {\bibinfo {author} {\bibfnamefont {P.}~\bibnamefont
  {Sekatski}}, \bibinfo {author} {\bibfnamefont {N.}~\bibnamefont {Sangouard}},
  \bibinfo {author} {\bibfnamefont {M.}~\bibnamefont {Stobińska}}, \bibinfo
  {author} {\bibfnamefont {F.}~\bibnamefont {Bussières}}, \bibinfo {author}
  {\bibfnamefont {M.}~\bibnamefont {Afzelius}}, \ and\ \bibinfo {author}
  {\bibfnamefont {N.}~\bibnamefont {Gisin}},\ }\href@noop {} {\bibfield
  {journal} {\bibinfo  {journal} {Physical Review A}\ }\textbf {\bibinfo
  {volume} {86}},\ \bibinfo {pages} {060301} (\bibinfo {year}
  {2012})}\BibitemShut {NoStop}%
\bibitem [{\citenamefont {Ghobadi}\ \emph {et~al.}(2013)\citenamefont
  {Ghobadi}, \citenamefont {Lvovsky},\ and\ \citenamefont
  {Simon}}]{Ghobadi2013}%
  \BibitemOpen
  \bibfield  {author} {\bibinfo {author} {\bibfnamefont {R.}~\bibnamefont
  {Ghobadi}}, \bibinfo {author} {\bibfnamefont {A.}~\bibnamefont {Lvovsky}}, \
  and\ \bibinfo {author} {\bibfnamefont {C.}~\bibnamefont {Simon}},\
  }\href@noop {} {\bibfield  {journal} {\bibinfo  {journal} {Physical Review
  Letters}\ }\textbf {\bibinfo {volume} {110}},\ \bibinfo {pages} {170406}
  (\bibinfo {year} {2013})}\BibitemShut {NoStop}%
\bibitem [{\citenamefont {Brune}\ \emph {et~al.}(1996)\citenamefont {Brune},
  \citenamefont {Hagley}, \citenamefont {Dreyer}, \citenamefont {Maître},
  \citenamefont {Maali}, \citenamefont {Wunderlich}, \citenamefont {Raimond},\
  and\ \citenamefont {Haroche}}]{Brune1996}%
  \BibitemOpen
  \bibfield  {author} {\bibinfo {author} {\bibfnamefont {M.}~\bibnamefont
  {Brune}}, \bibinfo {author} {\bibfnamefont {E.}~\bibnamefont {Hagley}},
  \bibinfo {author} {\bibfnamefont {J.}~\bibnamefont {Dreyer}}, \bibinfo
  {author} {\bibfnamefont {X.}~\bibnamefont {Maître}}, \bibinfo {author}
  {\bibfnamefont {A.}~\bibnamefont {Maali}}, \bibinfo {author} {\bibfnamefont
  {C.}~\bibnamefont {Wunderlich}}, \bibinfo {author} {\bibfnamefont {J.~M.}\
  \bibnamefont {Raimond}}, \ and\ \bibinfo {author} {\bibfnamefont
  {S.}~\bibnamefont {Haroche}},\ }\href@noop {} {\bibfield  {journal} {\bibinfo
   {journal} {Physical Review Letters}\ }\textbf {\bibinfo {volume} {77}},\
  \bibinfo {pages} {4887} (\bibinfo {year} {1996})}\BibitemShut {NoStop}%
\bibitem [{\citenamefont {Deleglise}\ \emph {et~al.}(2008)\citenamefont
  {Deleglise}, \citenamefont {Dotsenko}, \citenamefont {Sayrin}, \citenamefont
  {Bernu}, \citenamefont {Brune}, \citenamefont {Raimond},\ and\ \citenamefont
  {Haroche}}]{Deleglise2008}%
  \BibitemOpen
  \bibfield  {author} {\bibinfo {author} {\bibfnamefont {S.}~\bibnamefont
  {Deleglise}}, \bibinfo {author} {\bibfnamefont {I.}~\bibnamefont {Dotsenko}},
  \bibinfo {author} {\bibfnamefont {C.}~\bibnamefont {Sayrin}}, \bibinfo
  {author} {\bibfnamefont {J.}~\bibnamefont {Bernu}}, \bibinfo {author}
  {\bibfnamefont {M.}~\bibnamefont {Brune}}, \bibinfo {author} {\bibfnamefont
  {J.-M.}\ \bibnamefont {Raimond}}, \ and\ \bibinfo {author} {\bibfnamefont
  {S.}~\bibnamefont {Haroche}},\ }\href@noop {} {\bibfield  {journal} {\bibinfo
   {journal} {Nature}\ }\textbf {\bibinfo {volume} {455}},\ \bibinfo {pages}
  {510} (\bibinfo {year} {2008})}\BibitemShut {NoStop}%
\bibitem [{\citenamefont {Hofheinz}\ \emph {et~al.}(2009)\citenamefont
  {Hofheinz}, \citenamefont {Wang}, \citenamefont {Ansmann}, \citenamefont
  {Bialczak}, \citenamefont {Lucero}, \citenamefont {Neeley}, \citenamefont
  {O'Connell}, \citenamefont {Sank}, \citenamefont {Wenner}, \citenamefont
  {Martinis},\ and\ \citenamefont {Cleland}}]{Hofheinz2009}%
  \BibitemOpen
  \bibfield  {author} {\bibinfo {author} {\bibfnamefont {M.}~\bibnamefont
  {Hofheinz}}, \bibinfo {author} {\bibfnamefont {H.}~\bibnamefont {Wang}},
  \bibinfo {author} {\bibfnamefont {M.}~\bibnamefont {Ansmann}}, \bibinfo
  {author} {\bibfnamefont {R.~C.}\ \bibnamefont {Bialczak}}, \bibinfo {author}
  {\bibfnamefont {E.}~\bibnamefont {Lucero}}, \bibinfo {author} {\bibfnamefont
  {M.}~\bibnamefont {Neeley}}, \bibinfo {author} {\bibfnamefont {A.~D.}\
  \bibnamefont {O'Connell}}, \bibinfo {author} {\bibfnamefont {D.}~\bibnamefont
  {Sank}}, \bibinfo {author} {\bibfnamefont {J.}~\bibnamefont {Wenner}},
  \bibinfo {author} {\bibfnamefont {J.~M.}\ \bibnamefont {Martinis}}, \ and\
  \bibinfo {author} {\bibfnamefont {A.~N.}\ \bibnamefont {Cleland}},\
  }\href@noop {} {\bibfield  {journal} {\bibinfo  {journal} {Nature}\ }\textbf
  {\bibinfo {volume} {459}},\ \bibinfo {pages} {546} (\bibinfo {year}
  {2009})}\BibitemShut {NoStop}%
\bibitem [{\citenamefont {Guerlin}\ \emph {et~al.}(2007)\citenamefont
  {Guerlin}, \citenamefont {Bernu}, \citenamefont {Deleglise}, \citenamefont
  {Sayrin}, \citenamefont {Gleyzes}, \citenamefont {Kuhr}, \citenamefont
  {Brune}, \citenamefont {Raimond},\ and\ \citenamefont
  {Haroche}}]{Guerlin2007}%
  \BibitemOpen
  \bibfield  {author} {\bibinfo {author} {\bibfnamefont {C.}~\bibnamefont
  {Guerlin}}, \bibinfo {author} {\bibfnamefont {J.}~\bibnamefont {Bernu}},
  \bibinfo {author} {\bibfnamefont {S.}~\bibnamefont {Deleglise}}, \bibinfo
  {author} {\bibfnamefont {C.}~\bibnamefont {Sayrin}}, \bibinfo {author}
  {\bibfnamefont {S.}~\bibnamefont {Gleyzes}}, \bibinfo {author} {\bibfnamefont
  {S.}~\bibnamefont {Kuhr}}, \bibinfo {author} {\bibfnamefont {M.}~\bibnamefont
  {Brune}}, \bibinfo {author} {\bibfnamefont {J.~M.}\ \bibnamefont {Raimond}},
  \ and\ \bibinfo {author} {\bibfnamefont {S.}~\bibnamefont {Haroche}},\
  }\href@noop {} {\bibfield  {journal} {\bibinfo  {journal} {Nature}\ }\textbf
  {\bibinfo {volume} {448}},\ \bibinfo {pages} {889} (\bibinfo {year}
  {2007})}\BibitemShut {NoStop}%
\bibitem [{\citenamefont {Sayrin}\ \emph {et~al.}(2011)\citenamefont {Sayrin},
  \citenamefont {Dotsenko}, \citenamefont {Zhou}, \citenamefont {Peaudecerf},
  \citenamefont {Rybarczyk}, \citenamefont {Gleyzes}, \citenamefont {Rouchon},
  \citenamefont {Mirrahimi}, \citenamefont {Amini}, \citenamefont {Brune},
  \citenamefont {Raimond},\ and\ \citenamefont {Haroche}}]{Sayrin2011}%
  \BibitemOpen
  \bibfield  {author} {\bibinfo {author} {\bibfnamefont {C.}~\bibnamefont
  {Sayrin}}, \bibinfo {author} {\bibfnamefont {I.}~\bibnamefont {Dotsenko}},
  \bibinfo {author} {\bibfnamefont {X.}~\bibnamefont {Zhou}}, \bibinfo {author}
  {\bibfnamefont {B.}~\bibnamefont {Peaudecerf}}, \bibinfo {author}
  {\bibfnamefont {T.}~\bibnamefont {Rybarczyk}}, \bibinfo {author}
  {\bibfnamefont {S.}~\bibnamefont {Gleyzes}}, \bibinfo {author} {\bibfnamefont
  {P.}~\bibnamefont {Rouchon}}, \bibinfo {author} {\bibfnamefont
  {M.}~\bibnamefont {Mirrahimi}}, \bibinfo {author} {\bibfnamefont
  {H.}~\bibnamefont {Amini}}, \bibinfo {author} {\bibfnamefont
  {M.}~\bibnamefont {Brune}}, \bibinfo {author} {\bibfnamefont {J.-M.}\
  \bibnamefont {Raimond}}, \ and\ \bibinfo {author} {\bibfnamefont
  {S.}~\bibnamefont {Haroche}},\ }\href@noop {} {\bibfield  {journal} {\bibinfo
   {journal} {Nature}\ }\textbf {\bibinfo {volume} {477}},\ \bibinfo {pages}
  {73} (\bibinfo {year} {2011})}\BibitemShut {NoStop}%
\bibitem [{\citenamefont {Hofheinz}\ \emph {et~al.}(2008)\citenamefont
  {Hofheinz}, \citenamefont {Weig}, \citenamefont {Ansmann}, \citenamefont
  {Bialczak}, \citenamefont {Lucero}, \citenamefont {Neeley}, \citenamefont
  {O'Connell}, \citenamefont {Wang}, \citenamefont {Martinis},\ and\
  \citenamefont {Cleland}}]{Hofheinz2008}%
  \BibitemOpen
  \bibfield  {author} {\bibinfo {author} {\bibfnamefont {M.}~\bibnamefont
  {Hofheinz}}, \bibinfo {author} {\bibfnamefont {E.~M.}\ \bibnamefont {Weig}},
  \bibinfo {author} {\bibfnamefont {M.}~\bibnamefont {Ansmann}}, \bibinfo
  {author} {\bibfnamefont {R.~C.}\ \bibnamefont {Bialczak}}, \bibinfo {author}
  {\bibfnamefont {E.}~\bibnamefont {Lucero}}, \bibinfo {author} {\bibfnamefont
  {M.}~\bibnamefont {Neeley}}, \bibinfo {author} {\bibfnamefont {A.~D.}\
  \bibnamefont {O'Connell}}, \bibinfo {author} {\bibfnamefont {H.}~\bibnamefont
  {Wang}}, \bibinfo {author} {\bibfnamefont {J.~M.}\ \bibnamefont {Martinis}},
  \ and\ \bibinfo {author} {\bibfnamefont {A.~N.}\ \bibnamefont {Cleland}},\
  }\href@noop {} {\bibfield  {journal} {\bibinfo  {journal} {Nature}\ }\textbf
  {\bibinfo {volume} {454}},\ \bibinfo {pages} {310} (\bibinfo {year}
  {2008})}\BibitemShut {NoStop}%
\bibitem [{\citenamefont {Wang}\ \emph {et~al.}(2008)\citenamefont {Wang},
  \citenamefont {Hofheinz}, \citenamefont {Ansmann}, \citenamefont {Bialczak},
  \citenamefont {Lucero}, \citenamefont {Neeley}, \citenamefont {O’Connell},
  \citenamefont {Sank}, \citenamefont {Wenner}, \citenamefont {Cleland},\ and\
  \citenamefont {Martinis}}]{Wang2008}%
  \BibitemOpen
  \bibfield  {author} {\bibinfo {author} {\bibfnamefont {H.}~\bibnamefont
  {Wang}}, \bibinfo {author} {\bibfnamefont {M.}~\bibnamefont {Hofheinz}},
  \bibinfo {author} {\bibfnamefont {M.}~\bibnamefont {Ansmann}}, \bibinfo
  {author} {\bibfnamefont {R.~C.}\ \bibnamefont {Bialczak}}, \bibinfo {author}
  {\bibfnamefont {E.}~\bibnamefont {Lucero}}, \bibinfo {author} {\bibfnamefont
  {M.}~\bibnamefont {Neeley}}, \bibinfo {author} {\bibfnamefont {A.~D.}\
  \bibnamefont {O’Connell}}, \bibinfo {author} {\bibfnamefont
  {D.}~\bibnamefont {Sank}}, \bibinfo {author} {\bibfnamefont {J.}~\bibnamefont
  {Wenner}}, \bibinfo {author} {\bibfnamefont {A.~N.}\ \bibnamefont {Cleland}},
  \ and\ \bibinfo {author} {\bibfnamefont {J.~M.}\ \bibnamefont {Martinis}},\
  }\href@noop {} {\bibfield  {journal} {\bibinfo  {journal} {Physical Review
  Letters}\ }\textbf {\bibinfo {volume} {101}},\ \bibinfo {pages} {240401}
  (\bibinfo {year} {2008})}\BibitemShut {NoStop}%
\bibitem [{\citenamefont {Wang}\ \emph {et~al.}(2011)\citenamefont {Wang},
  \citenamefont {Mariantoni}, \citenamefont {Bialczak}, \citenamefont
  {Lenander}, \citenamefont {Lucero}, \citenamefont {Neeley}, \citenamefont
  {O’Connell}, \citenamefont {Sank}, \citenamefont {Weides}, \citenamefont
  {Wenner}, \citenamefont {Yamamoto}, \citenamefont {Yin}, \citenamefont
  {Zhao}, \citenamefont {Martinis},\ and\ \citenamefont {Cleland}}]{Wang2011}%
  \BibitemOpen
  \bibfield  {author} {\bibinfo {author} {\bibfnamefont {H.}~\bibnamefont
  {Wang}}, \bibinfo {author} {\bibfnamefont {M.}~\bibnamefont {Mariantoni}},
  \bibinfo {author} {\bibfnamefont {R.~C.}\ \bibnamefont {Bialczak}}, \bibinfo
  {author} {\bibfnamefont {M.}~\bibnamefont {Lenander}}, \bibinfo {author}
  {\bibfnamefont {E.}~\bibnamefont {Lucero}}, \bibinfo {author} {\bibfnamefont
  {M.}~\bibnamefont {Neeley}}, \bibinfo {author} {\bibfnamefont {A.~D.}\
  \bibnamefont {O’Connell}}, \bibinfo {author} {\bibfnamefont
  {D.}~\bibnamefont {Sank}}, \bibinfo {author} {\bibfnamefont {M.}~\bibnamefont
  {Weides}}, \bibinfo {author} {\bibfnamefont {J.}~\bibnamefont {Wenner}},
  \bibinfo {author} {\bibfnamefont {T.}~\bibnamefont {Yamamoto}}, \bibinfo
  {author} {\bibfnamefont {Y.}~\bibnamefont {Yin}}, \bibinfo {author}
  {\bibfnamefont {J.}~\bibnamefont {Zhao}}, \bibinfo {author} {\bibfnamefont
  {J.~M.}\ \bibnamefont {Martinis}}, \ and\ \bibinfo {author} {\bibfnamefont
  {A.~N.}\ \bibnamefont {Cleland}},\ }\href@noop {} {\bibfield  {journal}
  {\bibinfo  {journal} {Physical Review Letters}\ }\textbf {\bibinfo {volume}
  {106}},\ \bibinfo {pages} {060401} (\bibinfo {year} {2011})}\BibitemShut
  {NoStop}%
\bibitem [{\citenamefont {Koch}\ \emph {et~al.}(2010)\citenamefont {Koch},
  \citenamefont {Houck}, \citenamefont {Hur},\ and\ \citenamefont
  {Girvin}}]{Koch2010}%
  \BibitemOpen
  \bibfield  {author} {\bibinfo {author} {\bibfnamefont {J.}~\bibnamefont
  {Koch}}, \bibinfo {author} {\bibfnamefont {A.~A.}\ \bibnamefont {Houck}},
  \bibinfo {author} {\bibfnamefont {K.~L.}\ \bibnamefont {Hur}}, \ and\
  \bibinfo {author} {\bibfnamefont {S.~M.}\ \bibnamefont {Girvin}},\
  }\href@noop {} {\bibfield  {journal} {\bibinfo  {journal} {Physical Review
  A}\ }\textbf {\bibinfo {volume} {82}},\ \bibinfo {pages} {043811} (\bibinfo
  {year} {2010})}\BibitemShut {NoStop}%
\bibitem [{\citenamefont {Sliwa}\ \emph {et~al.}(2015)\citenamefont {Sliwa},
  \citenamefont {Hatridge}, \citenamefont {Narla}, \citenamefont {Shankar},
  \citenamefont {Frunzio}, \citenamefont {Schoelkopf},\ and\ \citenamefont
  {Devoret}}]{Sliwa2015}%
  \BibitemOpen
  \bibfield  {author} {\bibinfo {author} {\bibfnamefont {K.~M.}\ \bibnamefont
  {Sliwa}}, \bibinfo {author} {\bibfnamefont {M.}~\bibnamefont {Hatridge}},
  \bibinfo {author} {\bibfnamefont {A.}~\bibnamefont {Narla}}, \bibinfo
  {author} {\bibfnamefont {S.}~\bibnamefont {Shankar}}, \bibinfo {author}
  {\bibfnamefont {L.}~\bibnamefont {Frunzio}}, \bibinfo {author} {\bibfnamefont
  {R.~J.}\ \bibnamefont {Schoelkopf}}, \ and\ \bibinfo {author} {\bibfnamefont
  {M.~H.}\ \bibnamefont {Devoret}},\ }\href@noop {} {\bibfield  {journal}
  {\bibinfo  {journal} {Physical Review X}\ }\textbf {\bibinfo {volume} {5}},\
  \bibinfo {pages} {041020} (\bibinfo {year} {2015})}\BibitemShut {NoStop}%
\bibitem [{\citenamefont {Estep}\ \emph {et~al.}(2014)\citenamefont {Estep},
  \citenamefont {Sounas}, \citenamefont {Soric},\ and\ \citenamefont
  {Alu}}]{Estep2014}%
  \BibitemOpen
  \bibfield  {author} {\bibinfo {author} {\bibfnamefont {N.~A.}\ \bibnamefont
  {Estep}}, \bibinfo {author} {\bibfnamefont {D.~L.}\ \bibnamefont {Sounas}},
  \bibinfo {author} {\bibfnamefont {J.}~\bibnamefont {Soric}}, \ and\ \bibinfo
  {author} {\bibfnamefont {A.}~\bibnamefont {Alu}},\ }\href@noop {} {\bibfield
  {journal} {\bibinfo  {journal} {Nature Physics}\ }\textbf {\bibinfo {volume}
  {10}},\ \bibinfo {pages} {923} (\bibinfo {year} {2014})}\BibitemShut
  {NoStop}%
\bibitem [{\citenamefont {Goldman}\ and\ \citenamefont
  {Dalibard}(2014)}]{Goldman2014}%
  \BibitemOpen
  \bibfield  {author} {\bibinfo {author} {\bibfnamefont {N.}~\bibnamefont
  {Goldman}}\ and\ \bibinfo {author} {\bibfnamefont {J.}~\bibnamefont
  {Dalibard}},\ }\href@noop {} {\bibfield  {journal} {\bibinfo  {journal}
  {Physical Review X}\ }\textbf {\bibinfo {volume} {4}},\ \bibinfo {pages}
  {031027} (\bibinfo {year} {2014})}\BibitemShut {NoStop}%
\bibitem [{\citenamefont {Jotzu}\ \emph {et~al.}(2014)\citenamefont {Jotzu},
  \citenamefont {Messer}, \citenamefont {Desbuquois}, \citenamefont {Lebrat},
  \citenamefont {Uehlinger}, \citenamefont {Greif},\ and\ \citenamefont
  {Esslinger}}]{Jotzu2014a}%
  \BibitemOpen
  \bibfield  {author} {\bibinfo {author} {\bibfnamefont {G.}~\bibnamefont
  {Jotzu}}, \bibinfo {author} {\bibfnamefont {M.}~\bibnamefont {Messer}},
  \bibinfo {author} {\bibfnamefont {R.}~\bibnamefont {Desbuquois}}, \bibinfo
  {author} {\bibfnamefont {M.}~\bibnamefont {Lebrat}}, \bibinfo {author}
  {\bibfnamefont {T.}~\bibnamefont {Uehlinger}}, \bibinfo {author}
  {\bibfnamefont {D.}~\bibnamefont {Greif}}, \ and\ \bibinfo {author}
  {\bibfnamefont {T.}~\bibnamefont {Esslinger}},\ }\href@noop {} {\bibfield
  {journal} {\bibinfo  {journal} {Nature}\ }\textbf {\bibinfo {volume} {515}},\
  \bibinfo {pages} {237} (\bibinfo {year} {2014})}\BibitemShut {NoStop}%
\bibitem [{Sup()}]{Supp}%
  \BibitemOpen
  \href@noop {} {\bibinfo  {journal} {See Supplemental Material at
  http://link.aps.org/supplemental/10.1103/PhysRevLett.
  116.220502 for effective
  Hamiltonian, lattice dynamics, comparison with other schemes, cat state
  preparation and density matrix treatment of noises}\ }\BibitemShut {NoStop}%
\bibitem [{\citenamefont {Liberato}\ \emph {et~al.}(2007)\citenamefont
  {Liberato}, \citenamefont {Ciuti},\ and\ \citenamefont
  {Carusotto}}]{Liberato2007}%
  \BibitemOpen
\bibfield  {journal} {  }\bibfield  {author} {\bibinfo {author} {\bibfnamefont
  {S.~D.}\ \bibnamefont {Liberato}}, \bibinfo {author} {\bibfnamefont
  {C.}~\bibnamefont {Ciuti}}, \ and\ \bibinfo {author} {\bibfnamefont
  {I.}~\bibnamefont {Carusotto}},\ }\href@noop {} {\bibfield  {journal}
  {\bibinfo  {journal} {Physical Review Letters}\ }\textbf {\bibinfo {volume}
  {98}},\ \bibinfo {pages} {103602} (\bibinfo {year} {2007})}\BibitemShut
  {NoStop}%
\bibitem [{\citenamefont {Allman}\ \emph {et~al.}(2014)\citenamefont {Allman},
  \citenamefont {Whittaker}, \citenamefont {Castellanos-Beltran}, \citenamefont
  {Cicak}, \citenamefont {da~Silva}, \citenamefont {DeFeo}, \citenamefont
  {Lecocq}, \citenamefont {Sirois}, \citenamefont {Teufel}, \citenamefont
  {Aumentado},\ and\ \citenamefont {Simmonds}}]{Allman2014}%
  \BibitemOpen
  \bibfield  {author} {\bibinfo {author} {\bibfnamefont {M.~S.}\ \bibnamefont
  {Allman}}, \bibinfo {author} {\bibfnamefont {J.~D.}\ \bibnamefont
  {Whittaker}}, \bibinfo {author} {\bibfnamefont {M.}~\bibnamefont
  {Castellanos-Beltran}}, \bibinfo {author} {\bibfnamefont {K.}~\bibnamefont
  {Cicak}}, \bibinfo {author} {\bibfnamefont {F.}~\bibnamefont {da~Silva}},
  \bibinfo {author} {\bibfnamefont {M.~P.}\ \bibnamefont {DeFeo}}, \bibinfo
  {author} {\bibfnamefont {F.}~\bibnamefont {Lecocq}}, \bibinfo {author}
  {\bibfnamefont {A.}~\bibnamefont {Sirois}}, \bibinfo {author} {\bibfnamefont
  {J.~D.}\ \bibnamefont {Teufel}}, \bibinfo {author} {\bibfnamefont
  {J.}~\bibnamefont {Aumentado}}, \ and\ \bibinfo {author} {\bibfnamefont
  {R.~W.}\ \bibnamefont {Simmonds}},\ }\href@noop {} {\bibfield  {journal}
  {\bibinfo  {journal} {Physical Review Letters}\ }\textbf {\bibinfo {volume}
  {112}},\ \bibinfo {pages} {123601} (\bibinfo {year} {2014})}\BibitemShut
  {NoStop}%
\bibitem [{\citenamefont {Haldane}(1988)}]{Haldane1988}%
  \BibitemOpen
  \bibfield  {author} {\bibinfo {author} {\bibfnamefont {F.~D.~M.}\
  \bibnamefont {Haldane}},\ }\href@noop {} {\bibfield  {journal} {\bibinfo
  {journal} {Physical Review Letters}\ }\textbf {\bibinfo {volume} {61}},\
  \bibinfo {pages} {2015} (\bibinfo {year} {1988})}\BibitemShut {NoStop}%
\bibitem [{\citenamefont {Kane}\ and\ \citenamefont {Mele}(2005)}]{Kane2005}%
  \BibitemOpen
  \bibfield  {author} {\bibinfo {author} {\bibfnamefont {C.~L.}\ \bibnamefont
  {Kane}}\ and\ \bibinfo {author} {\bibfnamefont {E.~J.}\ \bibnamefont
  {Mele}},\ }\href@noop {} {\bibfield  {journal} {\bibinfo  {journal} {Physical
  Review Letters}\ }\textbf {\bibinfo {volume} {95}},\ \bibinfo {pages}
  {226801} (\bibinfo {year} {2005})}\BibitemShut {NoStop}%
\bibitem [{\citenamefont {Wang}\ \emph {et~al.}(2015)\citenamefont {Wang},
  \citenamefont {Cai}, \citenamefont {Yuan}, \citenamefont {Zhu},\ and\
  \citenamefont {Liu}}]{Wang2015b}%
  \BibitemOpen
  \bibfield  {author} {\bibinfo {author} {\bibfnamefont {D.-W.}\ \bibnamefont
  {Wang}}, \bibinfo {author} {\bibfnamefont {H.}~\bibnamefont {Cai}}, \bibinfo
  {author} {\bibfnamefont {L.}~\bibnamefont {Yuan}}, \bibinfo {author}
  {\bibfnamefont {S.-Y.}\ \bibnamefont {Zhu}}, \ and\ \bibinfo {author}
  {\bibfnamefont {R.-B.}\ \bibnamefont {Liu}},\ }\href@noop {} {\bibfield
  {journal} {\bibinfo  {journal} {Optica}\ }\textbf {\bibinfo {volume} {2}},\
  \bibinfo {pages} {712} (\bibinfo {year} {2015})}\BibitemShut {NoStop}%
\bibitem [{\citenamefont {Zou}\ \emph {et~al.}(2002)\citenamefont {Zou},
  \citenamefont {Pahlke},\ and\ \citenamefont {Mathis}}]{Zou2002}%
  \BibitemOpen
  \bibfield  {author} {\bibinfo {author} {\bibfnamefont {X.}~\bibnamefont
  {Zou}}, \bibinfo {author} {\bibfnamefont {K.}~\bibnamefont {Pahlke}}, \ and\
  \bibinfo {author} {\bibfnamefont {W.}~\bibnamefont {Mathis}},\ }\href@noop {}
  {\bibfield  {journal} {\bibinfo  {journal} {Physical Review A}\ }\textbf
  {\bibinfo {volume} {65}},\ \bibinfo {pages} {064303} (\bibinfo {year}
  {2002})}\BibitemShut {NoStop}%
\bibitem [{\citenamefont {Leuenberger}(2006)}]{Leuenberger2006}%
  \BibitemOpen
  \bibfield  {author} {\bibinfo {author} {\bibfnamefont {M.~N.}\ \bibnamefont
  {Leuenberger}},\ }\href@noop {} {\bibfield  {journal} {\bibinfo  {journal}
  {Physical Review B}\ }\textbf {\bibinfo {volume} {73}},\ \bibinfo {pages}
  {075312} (\bibinfo {year} {2006})}\BibitemShut {NoStop}%
\bibitem [{\citenamefont {Gerry}\ and\ \citenamefont
  {Campos}(2001)}]{Gerry2001}%
  \BibitemOpen
  \bibfield  {author} {\bibinfo {author} {\bibfnamefont {C.~C.}\ \bibnamefont
  {Gerry}}\ and\ \bibinfo {author} {\bibfnamefont {R.~A.}\ \bibnamefont
  {Campos}},\ }\href@noop {} {\bibfield  {journal} {\bibinfo  {journal}
  {Physical Review A}\ }\textbf {\bibinfo {volume} {64}},\ \bibinfo {pages}
  {063814} (\bibinfo {year} {2001})}\BibitemShut {NoStop}%
\bibitem [{\citenamefont {Sharypov}\ and\ \citenamefont
  {He}(2013)}]{Sharypov2013}%
  \BibitemOpen
  \bibfield  {author} {\bibinfo {author} {\bibfnamefont {A.~V.}\ \bibnamefont
  {Sharypov}}\ and\ \bibinfo {author} {\bibfnamefont {B.}~\bibnamefont {He}},\
  }\href@noop {} {\bibfield  {journal} {\bibinfo  {journal} {Physical Review
  A}\ }\textbf {\bibinfo {volume} {87}},\ \bibinfo {pages} {032323} (\bibinfo
  {year} {2013})}\BibitemShut {NoStop}%
\bibitem [{\citenamefont {Arnold}\ \emph {et~al.}(2015)\citenamefont {Arnold},
  \citenamefont {Demory}, \citenamefont {Loo}, \citenamefont {Lemaître},
  \citenamefont {Sagnes}, \citenamefont {Glazov}, \citenamefont {Krebs},
  \citenamefont {Voisin}, \citenamefont {Senellart},\ and\ \citenamefont
  {Lanco}}]{Arnold2015}%
  \BibitemOpen
  \bibfield  {author} {\bibinfo {author} {\bibfnamefont {C.}~\bibnamefont
  {Arnold}}, \bibinfo {author} {\bibfnamefont {J.}~\bibnamefont {Demory}},
  \bibinfo {author} {\bibfnamefont {V.}~\bibnamefont {Loo}}, \bibinfo {author}
  {\bibfnamefont {A.}~\bibnamefont {Lemaître}}, \bibinfo {author}
  {\bibfnamefont {I.}~\bibnamefont {Sagnes}}, \bibinfo {author} {\bibfnamefont
  {M.}~\bibnamefont {Glazov}}, \bibinfo {author} {\bibfnamefont
  {O.}~\bibnamefont {Krebs}}, \bibinfo {author} {\bibfnamefont
  {P.}~\bibnamefont {Voisin}}, \bibinfo {author} {\bibfnamefont
  {P.}~\bibnamefont {Senellart}}, \ and\ \bibinfo {author} {\bibfnamefont
  {L.}~\bibnamefont {Lanco}},\ }\href@noop {} {\bibfield  {journal} {\bibinfo
  {journal} {Nature Communications}\ }\textbf {\bibinfo {volume} {6}},\
  \bibinfo {pages} {6236} (\bibinfo {year} {2015})}\BibitemShut {NoStop}%
\bibitem [{\citenamefont {Krause}\ \emph {et~al.}(1987)\citenamefont {Krause},
  \citenamefont {Scully},\ and\ \citenamefont {Walther}}]{Krause1987}%
  \BibitemOpen
  \bibfield  {author} {\bibinfo {author} {\bibfnamefont {J.}~\bibnamefont
  {Krause}}, \bibinfo {author} {\bibfnamefont {M.~O.}\ \bibnamefont {Scully}},
  \ and\ \bibinfo {author} {\bibfnamefont {H.}~\bibnamefont {Walther}},\
  }\href@noop {} {\bibfield  {journal} {\bibinfo  {journal} {Physical Review
  A}\ }\textbf {\bibinfo {volume} {36}},\ \bibinfo {pages} {4547} (\bibinfo
  {year} {1987})}\BibitemShut {NoStop}%
\bibitem [{\citenamefont {Bertet}\ \emph {et~al.}(2002)\citenamefont {Bertet},
  \citenamefont {Osnaghi}, \citenamefont {Milman}, \citenamefont {Auffeves},
  \citenamefont {Maioli}, \citenamefont {Brune}, \citenamefont {Raimond},\ and\
  \citenamefont {Haroche}}]{Bertet2002}%
  \BibitemOpen
  \bibfield  {author} {\bibinfo {author} {\bibfnamefont {P.}~\bibnamefont
  {Bertet}}, \bibinfo {author} {\bibfnamefont {S.}~\bibnamefont {Osnaghi}},
  \bibinfo {author} {\bibfnamefont {P.}~\bibnamefont {Milman}}, \bibinfo
  {author} {\bibfnamefont {A.}~\bibnamefont {Auffeves}}, \bibinfo {author}
  {\bibfnamefont {P.}~\bibnamefont {Maioli}}, \bibinfo {author} {\bibfnamefont
  {M.}~\bibnamefont {Brune}}, \bibinfo {author} {\bibfnamefont {J.~M.}\
  \bibnamefont {Raimond}}, \ and\ \bibinfo {author} {\bibfnamefont
  {S.}~\bibnamefont {Haroche}},\ }\href@noop {} {\bibfield  {journal} {\bibinfo
   {journal} {Physical Review Letters}\ }\textbf {\bibinfo {volume} {88}},\
  \bibinfo {pages} {143601} (\bibinfo {year} {2002})}\BibitemShut {NoStop}%
\bibitem [{\citenamefont {Rigetti}\ \emph {et~al.}(2012)\citenamefont
  {Rigetti}, \citenamefont {Gambetta}, \citenamefont {Poletto}, \citenamefont
  {Plourde}, \citenamefont {Chow}, \citenamefont {Córcoles}, \citenamefont
  {Smolin}, \citenamefont {Merkel}, \citenamefont {Rozen}, \citenamefont
  {Keefe}, \citenamefont {Rothwell}, \citenamefont {Ketchen},\ and\
  \citenamefont {Steffen}}]{Rigetti2012}%
  \BibitemOpen
  \bibfield  {author} {\bibinfo {author} {\bibfnamefont {C.}~\bibnamefont
  {Rigetti}}, \bibinfo {author} {\bibfnamefont {J.~M.}\ \bibnamefont
  {Gambetta}}, \bibinfo {author} {\bibfnamefont {S.}~\bibnamefont {Poletto}},
  \bibinfo {author} {\bibfnamefont {B.~L.~T.}\ \bibnamefont {Plourde}},
  \bibinfo {author} {\bibfnamefont {J.~M.}\ \bibnamefont {Chow}}, \bibinfo
  {author} {\bibfnamefont {A.~D.}\ \bibnamefont {Córcoles}}, \bibinfo {author}
  {\bibfnamefont {J.~A.}\ \bibnamefont {Smolin}}, \bibinfo {author}
  {\bibfnamefont {S.~T.}\ \bibnamefont {Merkel}}, \bibinfo {author}
  {\bibfnamefont {J.~R.}\ \bibnamefont {Rozen}}, \bibinfo {author}
  {\bibfnamefont {G.~A.}\ \bibnamefont {Keefe}}, \bibinfo {author}
  {\bibfnamefont {M.~B.}\ \bibnamefont {Rothwell}}, \bibinfo {author}
  {\bibfnamefont {M.~B.}\ \bibnamefont {Ketchen}}, \ and\ \bibinfo {author}
  {\bibfnamefont {M.}~\bibnamefont {Steffen}},\ }\href@noop {} {\bibfield
  {journal} {\bibinfo  {journal} {Physical Review B}\ }\textbf {\bibinfo
  {volume} {86}},\ \bibinfo {pages} {100506} (\bibinfo {year}
  {2012})}\BibitemShut {NoStop}%
\bibitem [{\citenamefont {Jeffrey}\ \emph {et~al.}(2014)\citenamefont
  {Jeffrey}, \citenamefont {Sank}, \citenamefont {Mutus}, \citenamefont
  {White}, \citenamefont {Kelly}, \citenamefont {Barends}, \citenamefont
  {Chen}, \citenamefont {Chen}, \citenamefont {Chiaro}, \citenamefont
  {Dunsworth}, \citenamefont {Megrant}, \citenamefont {O’Malley},
  \citenamefont {Neill}, \citenamefont {Roushan}, \citenamefont {Vainsencher},
  \citenamefont {Wenner}, \citenamefont {Cleland},\ and\ \citenamefont
  {Martinis}}]{Jeffrey2014}%
  \BibitemOpen
  \bibfield  {author} {\bibinfo {author} {\bibfnamefont {E.}~\bibnamefont
  {Jeffrey}}, \bibinfo {author} {\bibfnamefont {D.}~\bibnamefont {Sank}},
  \bibinfo {author} {\bibfnamefont {J.~Y.}\ \bibnamefont {Mutus}}, \bibinfo
  {author} {\bibfnamefont {T.~C.}\ \bibnamefont {White}}, \bibinfo {author}
  {\bibfnamefont {J.}~\bibnamefont {Kelly}}, \bibinfo {author} {\bibfnamefont
  {R.}~\bibnamefont {Barends}}, \bibinfo {author} {\bibfnamefont
  {Y.}~\bibnamefont {Chen}}, \bibinfo {author} {\bibfnamefont {Z.}~\bibnamefont
  {Chen}}, \bibinfo {author} {\bibfnamefont {B.}~\bibnamefont {Chiaro}},
  \bibinfo {author} {\bibfnamefont {A.}~\bibnamefont {Dunsworth}}, \bibinfo
  {author} {\bibfnamefont {A.}~\bibnamefont {Megrant}}, \bibinfo {author}
  {\bibfnamefont {P.~J.~J.}\ \bibnamefont {O’Malley}}, \bibinfo {author}
  {\bibfnamefont {C.}~\bibnamefont {Neill}}, \bibinfo {author} {\bibfnamefont
  {P.}~\bibnamefont {Roushan}}, \bibinfo {author} {\bibfnamefont
  {A.}~\bibnamefont {Vainsencher}}, \bibinfo {author} {\bibfnamefont
  {J.}~\bibnamefont {Wenner}}, \bibinfo {author} {\bibfnamefont {A.~N.}\
  \bibnamefont {Cleland}}, \ and\ \bibinfo {author} {\bibfnamefont {J.~M.}\
  \bibnamefont {Martinis}},\ }\href@noop {} {\bibfield  {journal} {\bibinfo
  {journal} {Physical Review Letters}\ }\textbf {\bibinfo {volume} {112}},\
  \bibinfo {pages} {190504} (\bibinfo {year} {2014})}\BibitemShut {NoStop}%
\bibitem [{\citenamefont {Cleland}(2015)}]{Cleland2015}%
  \BibitemOpen
  \bibfield  {author} {\bibinfo {author} {\bibfnamefont {A.~N.}\ \bibnamefont
  {Cleland}},\ }\href@noop {} {\bibfield  {journal} {\bibinfo  {journal}
  {private communication}\ } (\bibinfo {year} {2015})}\BibitemShut {NoStop}%
\bibitem [{\citenamefont {Niemczyk}\ \emph {et~al.}(2010)\citenamefont
  {Niemczyk}, \citenamefont {Deppe}, \citenamefont {Huebl}, \citenamefont
  {Menzel}, \citenamefont {Hocke}, \citenamefont {Schwarz}, \citenamefont
  {Garcia-Ripoll}, \citenamefont {Zueco}, \citenamefont {Hummer}, \citenamefont
  {Solano}, \citenamefont {Marx},\ and\ \citenamefont {Gross}}]{Niemczyk2010}%
  \BibitemOpen
  \bibfield  {author} {\bibinfo {author} {\bibfnamefont {T.}~\bibnamefont
  {Niemczyk}}, \bibinfo {author} {\bibfnamefont {F.}~\bibnamefont {Deppe}},
  \bibinfo {author} {\bibfnamefont {H.}~\bibnamefont {Huebl}}, \bibinfo
  {author} {\bibfnamefont {E.~P.}\ \bibnamefont {Menzel}}, \bibinfo {author}
  {\bibfnamefont {F.}~\bibnamefont {Hocke}}, \bibinfo {author} {\bibfnamefont
  {M.~J.}\ \bibnamefont {Schwarz}}, \bibinfo {author} {\bibfnamefont {J.~J.}\
  \bibnamefont {Garcia-Ripoll}}, \bibinfo {author} {\bibfnamefont
  {D.}~\bibnamefont {Zueco}}, \bibinfo {author} {\bibfnamefont
  {T.}~\bibnamefont {Hummer}}, \bibinfo {author} {\bibfnamefont
  {E.}~\bibnamefont {Solano}}, \bibinfo {author} {\bibfnamefont
  {A.}~\bibnamefont {Marx}}, \ and\ \bibinfo {author} {\bibfnamefont
  {R.}~\bibnamefont {Gross}},\ }\href@noop {} {\bibfield  {journal} {\bibinfo
  {journal} {Nat Phys}\ }\textbf {\bibinfo {volume} {6}},\ \bibinfo {pages}
  {772} (\bibinfo {year} {2010})}\BibitemShut {NoStop}%
\bibitem [{\citenamefont {Forn-Diaz}\ \emph {et~al.}(2010)\citenamefont
  {Forn-Diaz}, \citenamefont {Lisenfeld}, \citenamefont {Marcos}, \citenamefont
  {Garcia-Ripoll}, \citenamefont {Solano}, \citenamefont {Harmans},\ and\
  \citenamefont {Mooij}}]{Forn2010}%
  \BibitemOpen
  \bibfield  {author} {\bibinfo {author} {\bibfnamefont {P.}~\bibnamefont
  {Forn-Diaz}}, \bibinfo {author} {\bibfnamefont {J.}~\bibnamefont
  {Lisenfeld}}, \bibinfo {author} {\bibfnamefont {D.}~\bibnamefont {Marcos}},
  \bibinfo {author} {\bibfnamefont {J.~J.}\ \bibnamefont {Garcia-Ripoll}},
  \bibinfo {author} {\bibfnamefont {E.}~\bibnamefont {Solano}}, \bibinfo
  {author} {\bibfnamefont {C.~J. P.~M.}\ \bibnamefont {Harmans}}, \ and\
  \bibinfo {author} {\bibfnamefont {J.~E.}\ \bibnamefont {Mooij}},\ }\href@noop
  {} {\bibfield  {journal} {\bibinfo  {journal} {Physical Review Letters}\
  }\textbf {\bibinfo {volume} {105}},\ \bibinfo {pages} {237001} (\bibinfo
  {year} {2010})}\BibitemShut {NoStop}%
\bibitem [{\citenamefont {{Reagor}}\ \emph {et~al.}(2015)\citenamefont
  {{Reagor}}, \citenamefont {{Pfaff}}, \citenamefont {{Axline}}, \citenamefont
  {{Heeres}}, \citenamefont {{Ofek}}, \citenamefont {{Sliwa}}, \citenamefont
  {{Holland}}, \citenamefont {{Wang}}, \citenamefont {{Blumoff}}, \citenamefont
  {{Chou}}, \citenamefont {{Hatridge}}, \citenamefont {{Frunzio}},
  \citenamefont {{Devoret}}, \citenamefont {{Jiang}},\ and\ \citenamefont
  {{Schoelkopf}}}]{Reagor2015}%
  \BibitemOpen
  \bibfield  {author} {\bibinfo {author} {\bibfnamefont {M.}~\bibnamefont
  {{Reagor}}}, \bibinfo {author} {\bibfnamefont {W.}~\bibnamefont {{Pfaff}}},
  \bibinfo {author} {\bibfnamefont {C.}~\bibnamefont {{Axline}}}, \bibinfo
  {author} {\bibfnamefont {R.~W.}\ \bibnamefont {{Heeres}}}, \bibinfo {author}
  {\bibfnamefont {N.}~\bibnamefont {{Ofek}}}, \bibinfo {author} {\bibfnamefont
  {K.}~\bibnamefont {{Sliwa}}}, \bibinfo {author} {\bibfnamefont
  {E.}~\bibnamefont {{Holland}}}, \bibinfo {author} {\bibfnamefont
  {C.}~\bibnamefont {{Wang}}}, \bibinfo {author} {\bibfnamefont
  {J.}~\bibnamefont {{Blumoff}}}, \bibinfo {author} {\bibfnamefont
  {K.}~\bibnamefont {{Chou}}}, \bibinfo {author} {\bibfnamefont {M.~J.}\
  \bibnamefont {{Hatridge}}}, \bibinfo {author} {\bibfnamefont
  {L.}~\bibnamefont {{Frunzio}}}, \bibinfo {author} {\bibfnamefont {M.~H.}\
  \bibnamefont {{Devoret}}}, \bibinfo {author} {\bibfnamefont {L.}~\bibnamefont
  {{Jiang}}}, \ and\ \bibinfo {author} {\bibfnamefont {R.~J.}\ \bibnamefont
  {{Schoelkopf}}},\ }\href@noop {} {\bibfield  {journal} {\bibinfo  {journal}
  {ArXiv e-prints}\ } (\bibinfo {year} {2015})},\ \Eprint
  {http://arxiv.org/abs/1508.05882} {arXiv:1508.05882 [quant-ph]} \BibitemShut
  {NoStop}%
\bibitem [{\citenamefont {Vissers}\ \emph {et~al.}(2015)\citenamefont
  {Vissers}, \citenamefont {Hubmayr}, \citenamefont {Sandberg}, \citenamefont
  {Chaudhuri}, \citenamefont {Bockstiegel},\ and\ \citenamefont
  {Gao}}]{Vissers2015}%
  \BibitemOpen
  \bibfield  {author} {\bibinfo {author} {\bibfnamefont {M.~R.}\ \bibnamefont
  {Vissers}}, \bibinfo {author} {\bibfnamefont {J.}~\bibnamefont {Hubmayr}},
  \bibinfo {author} {\bibfnamefont {M.}~\bibnamefont {Sandberg}}, \bibinfo
  {author} {\bibfnamefont {S.}~\bibnamefont {Chaudhuri}}, \bibinfo {author}
  {\bibfnamefont {C.}~\bibnamefont {Bockstiegel}}, \ and\ \bibinfo {author}
  {\bibfnamefont {J.}~\bibnamefont {Gao}},\ }\href@noop {} {\bibfield
  {journal} {\bibinfo  {journal} {Applied Physics Letters}\ }\textbf {\bibinfo
  {volume} {107}},\ \bibinfo {pages} {062601} (\bibinfo {year}
  {2015})}\BibitemShut {NoStop}%
\bibitem [{\citenamefont {Palacios-Laloy}\ \emph {et~al.}(2008)\citenamefont
  {Palacios-Laloy}, \citenamefont {Nguyen}, \citenamefont {Mallet},
  \citenamefont {Bertet}, \citenamefont {Vion},\ and\ \citenamefont
  {Esteve}}]{Laloy2008}%
  \BibitemOpen
  \bibfield  {author} {\bibinfo {author} {\bibfnamefont {A.}~\bibnamefont
  {Palacios-Laloy}}, \bibinfo {author} {\bibfnamefont {F.}~\bibnamefont
  {Nguyen}}, \bibinfo {author} {\bibfnamefont {F.}~\bibnamefont {Mallet}},
  \bibinfo {author} {\bibfnamefont {P.}~\bibnamefont {Bertet}}, \bibinfo
  {author} {\bibfnamefont {D.}~\bibnamefont {Vion}}, \ and\ \bibinfo {author}
  {\bibfnamefont {D.}~\bibnamefont {Esteve}},\ }\href@noop {} {\bibfield
  {journal} {\bibinfo  {journal} {Journal of Low Temperature Physics}\ }\textbf
  {\bibinfo {volume} {151}},\ \bibinfo {pages} {1034} (\bibinfo {year}
  {2008})}\BibitemShut {NoStop}%
\bibitem [{\citenamefont {Sandberg}\ \emph {et~al.}(2008)\citenamefont
  {Sandberg}, \citenamefont {Wilson}, \citenamefont {Persson}, \citenamefont
  {Bauch}, \citenamefont {Johansson}, \citenamefont {Shumeiko}, \citenamefont
  {Duty},\ and\ \citenamefont {Delsing}}]{Sandberg2008}%
  \BibitemOpen
  \bibfield  {author} {\bibinfo {author} {\bibfnamefont {M.}~\bibnamefont
  {Sandberg}}, \bibinfo {author} {\bibfnamefont {C.~M.}\ \bibnamefont
  {Wilson}}, \bibinfo {author} {\bibfnamefont {F.}~\bibnamefont {Persson}},
  \bibinfo {author} {\bibfnamefont {T.}~\bibnamefont {Bauch}}, \bibinfo
  {author} {\bibfnamefont {G.}~\bibnamefont {Johansson}}, \bibinfo {author}
  {\bibfnamefont {V.}~\bibnamefont {Shumeiko}}, \bibinfo {author}
  {\bibfnamefont {T.}~\bibnamefont {Duty}}, \ and\ \bibinfo {author}
  {\bibfnamefont {P.}~\bibnamefont {Delsing}},\ }\href@noop {} {\bibfield
  {journal} {\bibinfo  {journal} {Applied Physics Letters}\ }\textbf {\bibinfo
  {volume} {92}},\ \bibinfo {pages} {203501} (\bibinfo {year}
  {2008})}\BibitemShut {NoStop}%
\bibitem [{\citenamefont {Johansson}\ \emph {et~al.}(2009)\citenamefont
  {Johansson}, \citenamefont {Johansson}, \citenamefont {Wilson},\ and\
  \citenamefont {Nori}}]{Johansson2009}%
  \BibitemOpen
  \bibfield  {author} {\bibinfo {author} {\bibfnamefont {J.~R.}\ \bibnamefont
  {Johansson}}, \bibinfo {author} {\bibfnamefont {G.}~\bibnamefont
  {Johansson}}, \bibinfo {author} {\bibfnamefont {C.~M.}\ \bibnamefont
  {Wilson}}, \ and\ \bibinfo {author} {\bibfnamefont {F.}~\bibnamefont
  {Nori}},\ }\href@noop {} {\bibfield  {journal} {\bibinfo  {journal} {Physical
  Review Letters}\ }\textbf {\bibinfo {volume} {103}},\ \bibinfo {pages}
  {147003} (\bibinfo {year} {2009})}\BibitemShut {NoStop}%
\bibitem [{\citenamefont {Wilson}\ \emph {et~al.}(2011)\citenamefont {Wilson},
  \citenamefont {Johansson}, \citenamefont {Pourkabirian}, \citenamefont
  {Simoen}, \citenamefont {Johansson}, \citenamefont {Duty}, \citenamefont
  {Nori},\ and\ \citenamefont {Delsing}}]{Wilson2011}%
  \BibitemOpen
  \bibfield  {author} {\bibinfo {author} {\bibfnamefont {C.~M.}\ \bibnamefont
  {Wilson}}, \bibinfo {author} {\bibfnamefont {G.}~\bibnamefont {Johansson}},
  \bibinfo {author} {\bibfnamefont {A.}~\bibnamefont {Pourkabirian}}, \bibinfo
  {author} {\bibfnamefont {M.}~\bibnamefont {Simoen}}, \bibinfo {author}
  {\bibfnamefont {J.~R.}\ \bibnamefont {Johansson}}, \bibinfo {author}
  {\bibfnamefont {T.}~\bibnamefont {Duty}}, \bibinfo {author} {\bibfnamefont
  {F.}~\bibnamefont {Nori}}, \ and\ \bibinfo {author} {\bibfnamefont
  {P.}~\bibnamefont {Delsing}},\ }\href@noop {} {\bibfield  {journal} {\bibinfo
   {journal} {Nature}\ }\textbf {\bibinfo {volume} {479}},\ \bibinfo {pages}
  {376} (\bibinfo {year} {2011})}\BibitemShut {NoStop}%
\end{thebibliography}%

\end{document}